\documentclass{article}

\usepackage{arxiv}

\usepackage[utf8]{inputenc} 
\usepackage[T1]{fontenc}    
\usepackage{hyperref}       
\usepackage{url}            
\usepackage{booktabs}       
\usepackage{amsfonts}       
\usepackage{nicefrac}       
\usepackage{microtype}      
\usepackage{graphicx}
\usepackage{subcaption}
\usepackage{xcolor}

\newcommand{\demandsatisfaction}{d}
\newcommand{\networkcost}{c}
\newcommand{\robustness}{r}
\newcommand{\zsc}{\sigma}

\title{Local topological features of robust supply networks}

\author{
  Alexey Lyutov\\
  Department of Mathematics \& Logistics\\
  Jacobs University Bremen gGmbH\\
  Campus Ring 1\\
  28759 Bremen, Germany\\
  \texttt{a.lyutov@jacobs-university.de} 
  \And
  Yilmaz Uygun\\
  Department of Mathematics \& Logistics\\
  Jacobs University Bremen gGmbH\\
  Campus Ring 1\\
  28759 Bremen, Germany\\
  \texttt{y.uygun@jacobs-university.de} \\
   \And
  Marc-Thorsten Hütt\\
  Department of Life Sciences \& Chemistry\\
  Jacobs University Bremen gGmbH\\
  Campus Ring 1\\
  28759 Bremen, Germany\\
  \texttt{m.huett@jacobs-university.de} \\
\\
}

\begin{document}
\maketitle

\begin{abstract}
The design of robust supply and distribution systems is one of the fundamental challenges at the interface of network science and logistics. Given the multitude of performance criteria, real-world constraints, and external influences acting upon such a system, even formulating an appropriate research question to address this topic is non-trivial. 

Here we present an abstraction of a supply and distribution system leading to a minimal model, which only retains stylized facts of the systemic function and, in this way, allows us to investigate the generic properties of robust supply networks. On this level of abstraction, a supply and distribution system is the strategic use of transportation to eliminate mismatches between production patterns (i.e., the amounts of goods produced at each production site of a company) and demand patterns (i.e., the amount of goods consumed at each location). When creating networks based on this paradigm and furthermore requiring the robustness of the system with respect to the loss of transportation routes (edge of the network) we see that robust networks are built from specific sets of subgraphs, while vulnerable networks display a markedly different subgraph composition. 

Our findings confirm a long-standing hypothesis in the field of network science, namely, that network \textit{motifs} -- statistically over-represented small subgraphs -- are informative about the robust functioning of a network. Also, our findings offer a blueprint for enhancing the robustness of real-world supply and distribution systems. 
    
\end{abstract}

\keywords{network motifs \and minimal model \and supply chain management \and spatial networks}

\section{Introduction}
\label{intro}
The analysis of network motifs \cite{shen2002network,Milo2004,alon2007network} goes back to the early phase of network science \cite{strogatz2001exploring,albert2002statistical,barabasi2016network}. In contrast to studying the large-scale topological features of complex networks (e.g., their broad degree distribution or their hierarchical organization) or the microscale of properties of individual nodes (e.g., the betweenness centrality or the local clustering coefficient), network motifs have drawn the attention to a 'mesoscale'\footnote{Note that the modular organization and community structure of complex networks is of course another mesoscale feature on a slightly larger level of organization, which revealed itself as highly relevant for the functional organization of complex networks \cite{newman2006modularity,guimera2005cartography,hutt2019modular}. Note that one can also quantitatively study the interplay of these two mesoscales in complex networks \cite{Fretter2012,Beber:2012ix}.} with the hope of explaining some of the functional properties of complex networks via the networks' non-random features on this scale of organization.

In fact, \textit{motif signatures} -- patterns of non-random occurrences of certain few-node subgraphs -- have been identified and associated to the networks' functional categories \cite{Milo2004} and have been, in subsequent studies, linked to the robustness of the networks' dynamical function, e.g., for Boolean dynamics \cite{klemm2005topology} and flow networks evolved towards a robust performance under random deletion of links or nodes \cite{kaluza2007design,kaluza_evolutionary_2007,kaluza2008self,beber2013prescribed}. 

The question addressed in our investigation is whether the deep relationship between motif signatures and robust functioning identified in network science translates to supply networks as well. 

For supply networks robust functioning is of utmost importance. Key aspects in the infrastructure of our industrialized world depend on it. As a consequence, the topic of supply network robustness has received substantial scientific attention.  Methods from nonlinear dynamics have been applied to study supply and distribution networks under fluctuations (e.g. \cite{ritterskamp2018revealing, demirel2019identifying}). 
Using methods of nonlinear dynamics, especially parameterizing fixed points together with a stochastic sampling of the unknown entries of the Jacobi matrix (generalized modeling, \cite{gross2006generalized}) were analyzed in \cite{ritterskamp2018revealing} and \cite{demirel2019identifying} (see also \cite{gross2018introduction}). 

At the same time, supply and distribution networks are high-dimensional systems with high demands on efficient organization and the fulfillment of logistic target values. These aspects are often addressed by optimization methods (e.g. \cite{hendriks2012design, garcia2015supply}).  In \cite{armbruster2011structural, hendriks2012design}, an abstract formulation of logistic networks (supply and distribution networks) has been formulated as an optimization problem. 

Methods from network science have been particularly employed to supply systems to analyze the impact of disruptions, such as transportation failure or supply shortages, and hence the robustness and resilience of such systems (e.g. \cite{helbing2004physics,sun2005scale,atalay2011network,brintrup2018supply}). In \cite{atalay2011network}, the value of network representations of supply networks for an understanding of economic processes was elaborated, with an application focus on the automotive industry. 

The resilience and vulnerability of supply chains and supply networks to disruption were analyzed -- particularly in light of the COVID-19 pandemic -- in \cite{ivanov2020viability} and \cite{ivanov2020viable}. An overview of the important field of mathematical modeling of sustainable supply chains is provided by \cite{seuring2013review}. 

The embedding of supply networks in real geographical space, the often multi-modal nature of supply networks (distributing not a single good or material, but rather a whole range of goods and materials, which are often interdependent), as well as the weighted nature of supply networks (where suitable weights of edges are the total \textit{volume} shipped along this edge in a certain time window, or the total \textit{value} or the average \textit{cost} per shipment, which in turn is partly related to geographical distance) all make formal network representations suitable for the analysis, e.g., of network motifs, challenging. 

In order to account for these incompatibilities between abstract network representations and real-world features of these systems, we introduce a stylized supply network model, which retains the spatial embedding and the overall 'source-to-target' organization of supply networks, but is generic enough to allow for a motif analysis of the resulting networks.

\newpage
\section{Methods}
\subsection{Supply network model}
\begin{figure*}[ht]
    \centering
    \begin{minipage}[t]{.4\linewidth}
        \centering
        \includegraphics[width=\columnwidth]{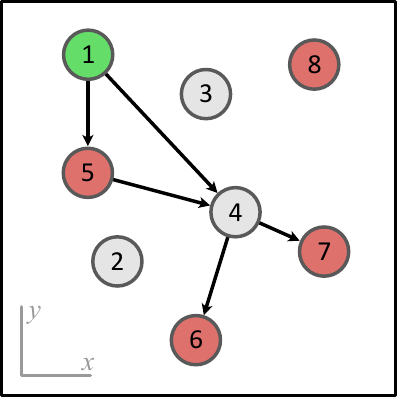}
        \subcaption{$\demandsatisfaction=0.75$, $\robustness=0.4$}
        \label{fig:model-f1}
    \end{minipage}
    \begin{minipage}[t]{.4\linewidth}
        \centering
        \includegraphics[width=\columnwidth]{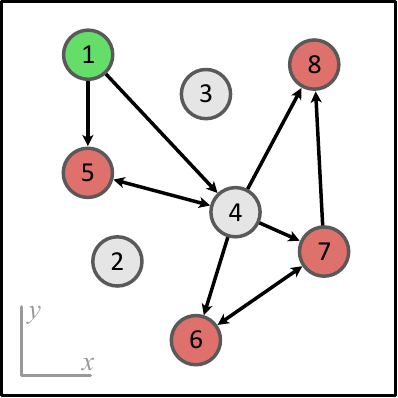}
        \subcaption{$\demandsatisfaction=1.0$, $\robustness=1.0$}
        \label{fig:model-f2}
    \end{minipage}
    \captionsetup{width=0.95\columnwidth}
    \caption{Examples of an average (a) and high (b) robustness network as an illustration of demand satisfaction $\demandsatisfaction$ and robustness $\robustness$ in our supply network model. Values for $\demandsatisfaction$ and $\robustness$ are indicated below each network.}
    \label{fig:model}
\end{figure*}

Our supply network model consists of $N$ nodes that are spatially distributed on a 2D plane (Fig.~\ref{fig:model}). Each node can have one of three roles: producers (green in Fig.~\ref{fig:model}) that are generating a product, demanders (red) that require the product to be delivered, and intermediate nodes (gray) that neither produce, nor demand the product, but can be used to deliver the product efficiently (warehouses). A \textbf{model setup} is a set of $N$ nodes, each with $(x,y)$ coordinates and an assigned role. All setups used in the experiments in this paper have only one producing node and $N_{d}=N/2$ demanders. The coordinates of nodes for a single setup are sampled from a uniform distribution. 

For a given setup there are multiple \textbf{supply networks} -- sets of directed edges, that connect the nodes. Each network has the following parameters: number of edges $M$, network cost $\networkcost$, demand satisfaction $\demandsatisfaction$, robustness $\robustness$. Network cost $\networkcost$ is the sum of the Euclidean distances of all edges in the network. It reflects how optimal the product paths in the network are with respect to edge length. Demand satisfaction $\demandsatisfaction$ is calculated as the percentage of satisfied demanders. A demander is satisfied if there exists a path from a producer to this demander. The demand satisfaction reflects how well the network performs in fulfilling an existing demand pattern. Robustness is a metric that shows how susceptible the network is to a random loss of edges. It is computed by finding the subset of edges $E_{\robustness}$ such that any edge from $E_{\robustness}$ can be safely removed from the initial network and the $\demandsatisfaction$ of the resulting network will not decrease. The robustness $\robustness$, in this case, is $\robustness=|E_{\robustness}|/M$. In all simulations in this paper the generated networks were required to have a full demand satisfaction, $\demandsatisfaction=1.0$.

A focus of our investigation is the analysis of these supply networks from the perspective of few-node subgraphs. To this end, we follow the concept of a \textit{motif signature} proposed by Milo et al.~\cite{Milo2004}. This signature shows the frequency of each of 13 connected 3-node subgraphs in the original network, compared to its randomized versions via a normalized vector of z-scores. Randomized networks can be generated using different versions of the null model. Here we have used the default setup used in the original research~\cite{Milo2004}, the version that does not preserve the mutual edges, and our custom null model that generates networks with the same level of demand satisfaction as the original network. To make the results more comparable, the default null model that preserves mutual edges has been used in the main part of our investigation, while the examples of results for the other null model variants are shown as Supplementary Information. In general, the association of robustness with a non-random subgraph composition is observed in all null model variants. Motif calculations were performed using the mfinder software developed by Alon et al.~\cite{Milo2002}.

\subsection{Numerical simulations}
To solve a given setup of the model, it is necessary to find a network that connects producers and demanders. Finding the solution that gives the lowest $\networkcost$ while having a full demand satisfaction is an NP-hard Metric Steiner Tree problem~\cite{Hwang1992}. A heuristic solution used in this paper is based on finding the exact solution for a subset of nodes that includes only producers and demanders and then sequentially adding intermediate nodes until the network cost stops decreasing. 

\begin{figure}[ht]
    \centering
    \includegraphics[width=.4\linewidth]{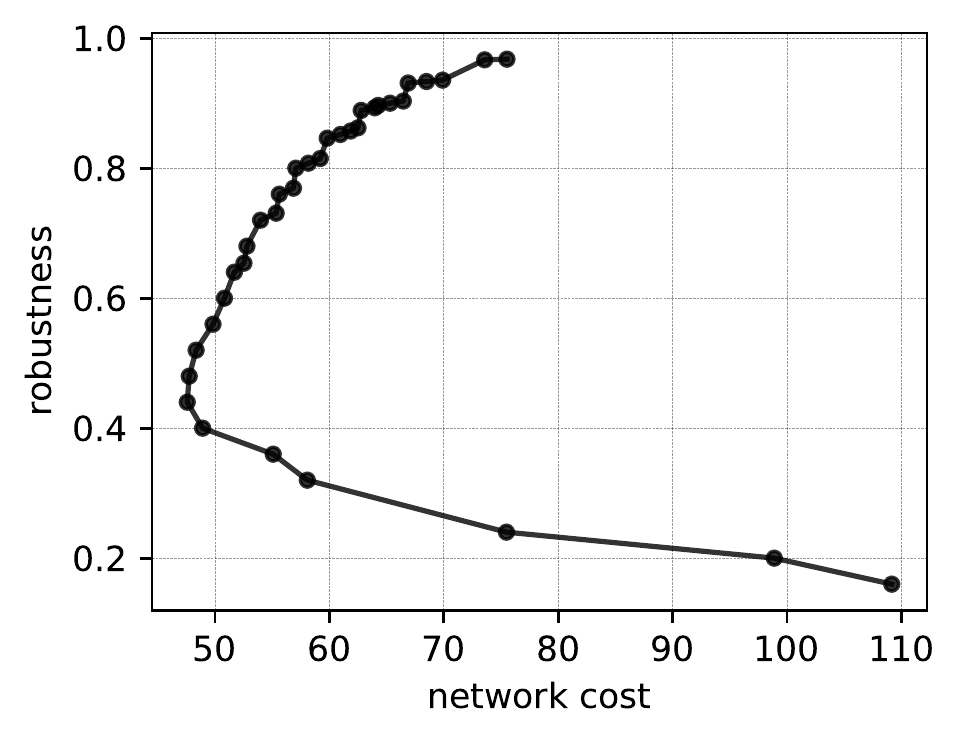}
    \caption{An example of a typical Pareto front of an ($\networkcost$, $\robustness$) optimization. }
    \label{fig:opt-front}
\end{figure}

The problem of generating robust networks is computationally more complex and has no analytical solution or simple yet efficient heuristic. It can be formulated as a multi-objective optimization problem with objective functions $min(\networkcost)$, $max(\robustness)$, and additional constraints, e.g. the minimal demand satisfaction $\demandsatisfaction$, the number of edges $M$, or further constraints on the edge lengths. The optimization problem is solved using genetic algorithms~\cite{Deb2002} with a small modification that allows simultaneous maximization and minimization of a target objective. This modification might be necessary in optimization problems where the allowed number of edges in a network has a lower boundary $M_0 \leq M$. The shape of the Pareto front in this case might have two parts: below and above the minimal network cost (see example in Fig.~\ref{fig:opt-front}). In this case, for the networks that have robustness lower than the robustness of the network with the best $\networkcost$ ($\robustness<0.44$ in the figure), the problem of minimizing the robustness is solved. For the networks above, the robustness is maximized. The critical robustness is re-evaluated after each generation, as the front evolves.

During the optimization process, new networks are either generated randomly -- by picking $M$ random edges or created by mutating and recombining optimal networks. There are three possible mutation procedures -- removing, adding, or replacing $M_m$ random edges. The recombination procedure takes two networks, selects a random subset of edges from each, and outputs the network with the union of the selected edges. The first generation consists of random networks only. Each following generation is composed of the 30\% best networks from the previous generation -- Pareto front and lowest-rank networks. The remaining 70\% of the generation are mutations and recombinations of these 30\% best and some random networks. The process is repeated for $G_{N}$ generations without an explicit convergence stop.

\section{Results}
\subsection{Robust supply networks}
In the first set of numerical experiments, we investigated how the robustness shapes supply networks on a structural level. To do so, we optimized robustness $\robustness$ and network cost $\networkcost$ for 50 setups with the different spatial distribution of nodes while keeping the node roles fixed. It is clear that the allowed number of edges in a network $M$ has a direct impact on the robustness. It is much easier to install robustness with more edges, as the number of alternative paths increases rapidly with higher connectivity, e.g. by having direct edges from the producing node and duplicating them through one intermediate node. On the other hand, when $M$ is close to the minimal number of edges required for full demand satisfaction, no edges can be used as alternative paths. To make networks in both cases more specific, we have defined the minimum $M_{min}$ and the maximum $M_{max}$ number of edges allowed in the optimal networks and varied these boundaries, solving 50 different setups for each pair ($M_{min}$, $M_{max}$). 

\begin{figure*}[ht]
    \centering
    \begin{minipage}[t]{.3\linewidth}
        \centering
        \includegraphics[height=\columnwidth]{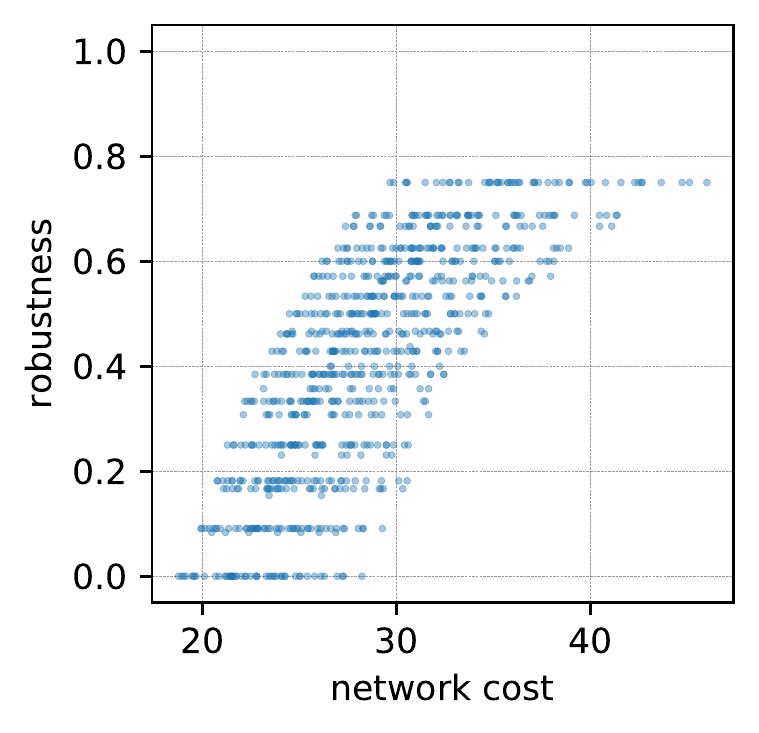}\\
	$M \in [10, 16]$
    \end{minipage}
    \begin{minipage}[t]{.3\linewidth}
        \centering
        \includegraphics[height=\columnwidth]{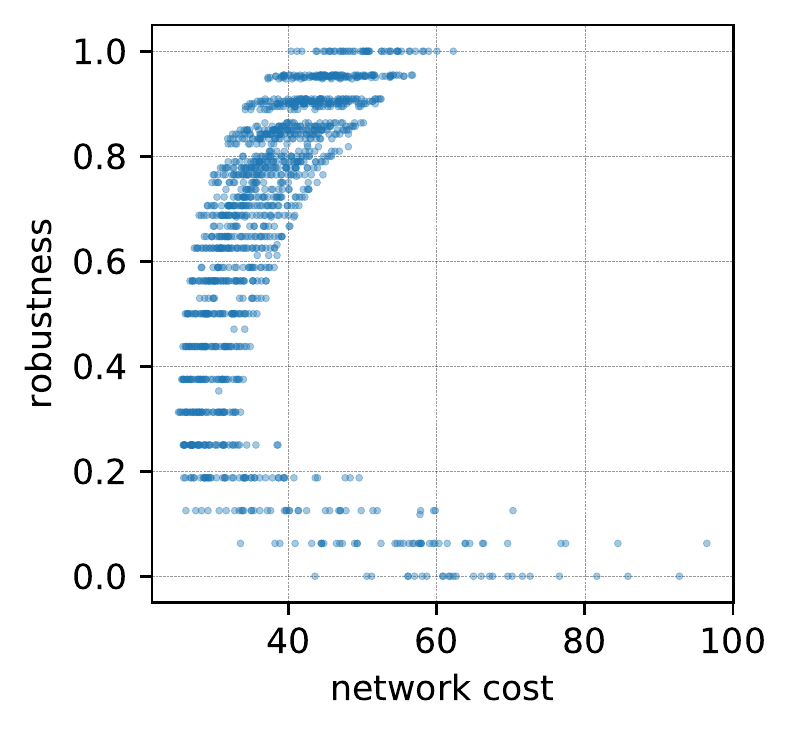}\\
	$M \in [16, 22]$
    \end{minipage}
    \begin{minipage}[t]{.3\linewidth}
        \centering
        \includegraphics[height=\columnwidth]{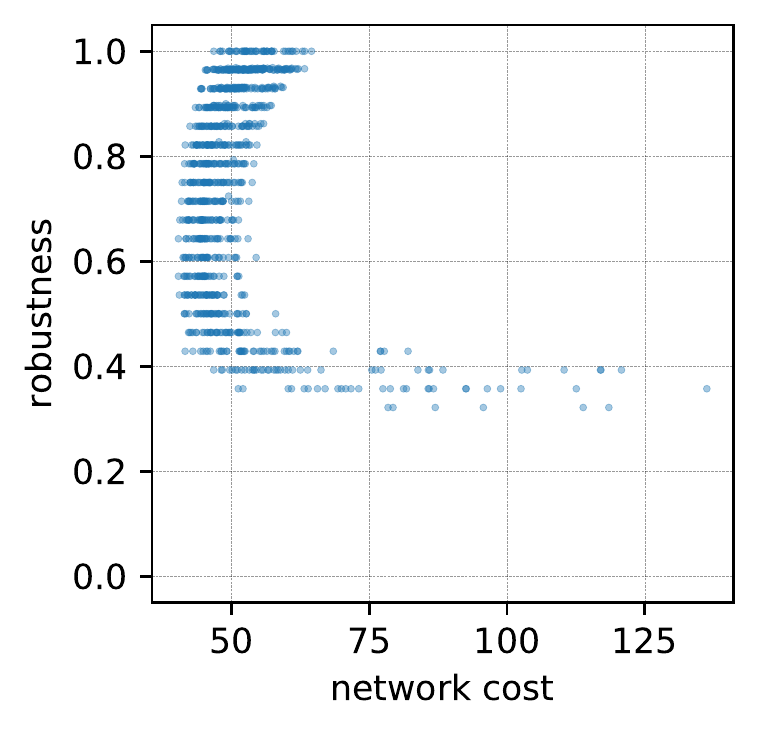}\\
	$M \in [28, 34]$
    \end{minipage}
    \captionsetup{width=0.90\columnwidth}
    \caption{Pareto fronts in a series of $(\networkcost, \robustness)$ optimizations with $N=20$ and varying $M$ boundaries. Each figure combines the results of 50 model setup runs that have different spatial distribution of nodes.}
    \label{fig:N=20-nc-r}
\end{figure*}

The first series of optimizations are performed for $N=20$ nodes with $(M_{min}, M_{max}) \in [(10, 16), (13, 19), ...,\allowbreak (40, 46)]$. Inspecting the optimization results in form of Pareto fronts (Fig.~\ref{fig:N=20-nc-r}) it can be seen that depending on the allowed number of edges high or low robustness areas become less populated. To analyze the structures of robust and vulnerable networks, we are sampling one network with low and one network with high robustness from each of the 50 optimizations. The low and high robustness, in this case, are defined as 10 and 90 percentiles of the $\robustness$ values for each edge limit. 

\begin{figure*}[h]
    \centering
    \begin{minipage}[t]{.45\linewidth}
        \centering
        \includegraphics[width=\columnwidth]{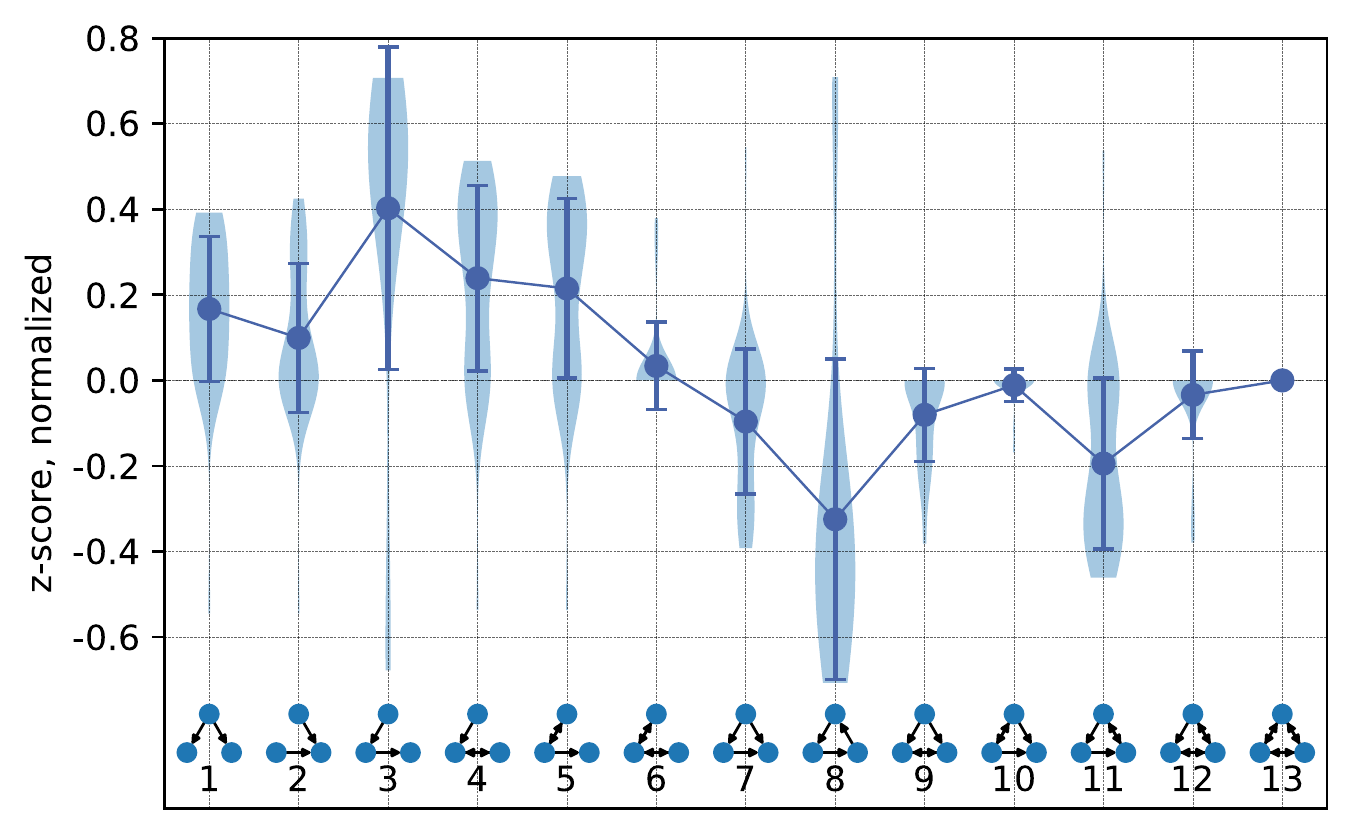}\\
	$M \in [16, 22]$, $\robustness=0.19$
    \end{minipage}
    \begin{minipage}[t]{.45\linewidth}
        \centering
        \includegraphics[width=\columnwidth]{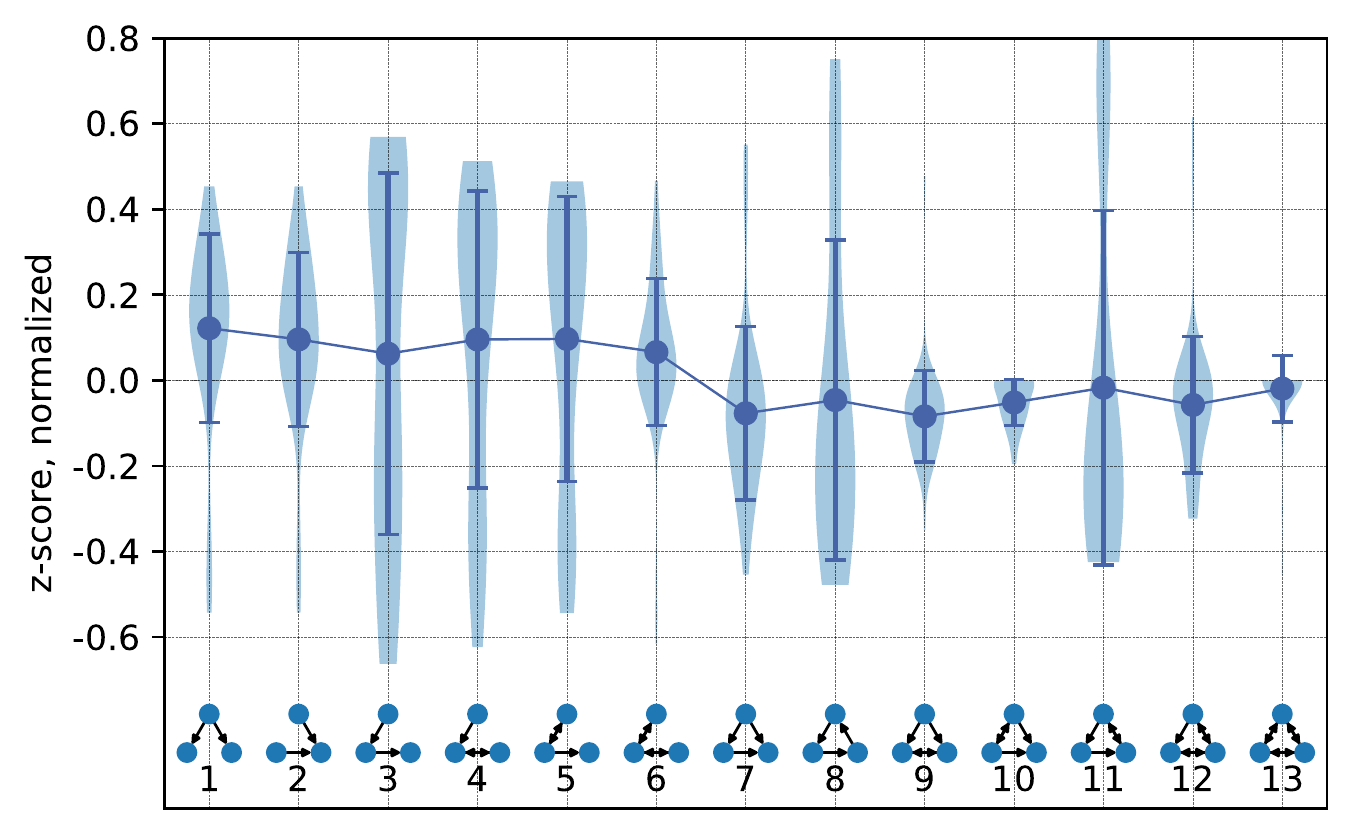}\\
	$M \in [22, 28]$, $\robustness=0.32$
    \end{minipage}
    \captionsetup{width=0.90\columnwidth}
    \caption{We generate 50 different network setups with $N=20$ and for each setup perform a $(\networkcost, \robustness)$ optimization with restriction on the allowed number of edges $M$ (shown below each figure). From the resulting Pareto fronts, we take 50 vulnerable networks with given $\robustness$ and compute their motifs. For a single network, the result of motif computations are 13 z-scores that indicate how over- or underrepresented each subgraph is in the original network, compared to its randomized versions. This gives 50 z-score values for each of the subgraphs that form a distribution drawn with shaded vertical violin plots. Blue circles in the figures are the mean values and vertical lines with ticks are the standard deviations of these distributions.}
    \label{fig:N=20-zscores-vulnerable}
\end{figure*}

\begin{figure*}[h]
    \centering
    \begin{minipage}[t]{.45\linewidth}
        \centering
        \includegraphics[width=\columnwidth]{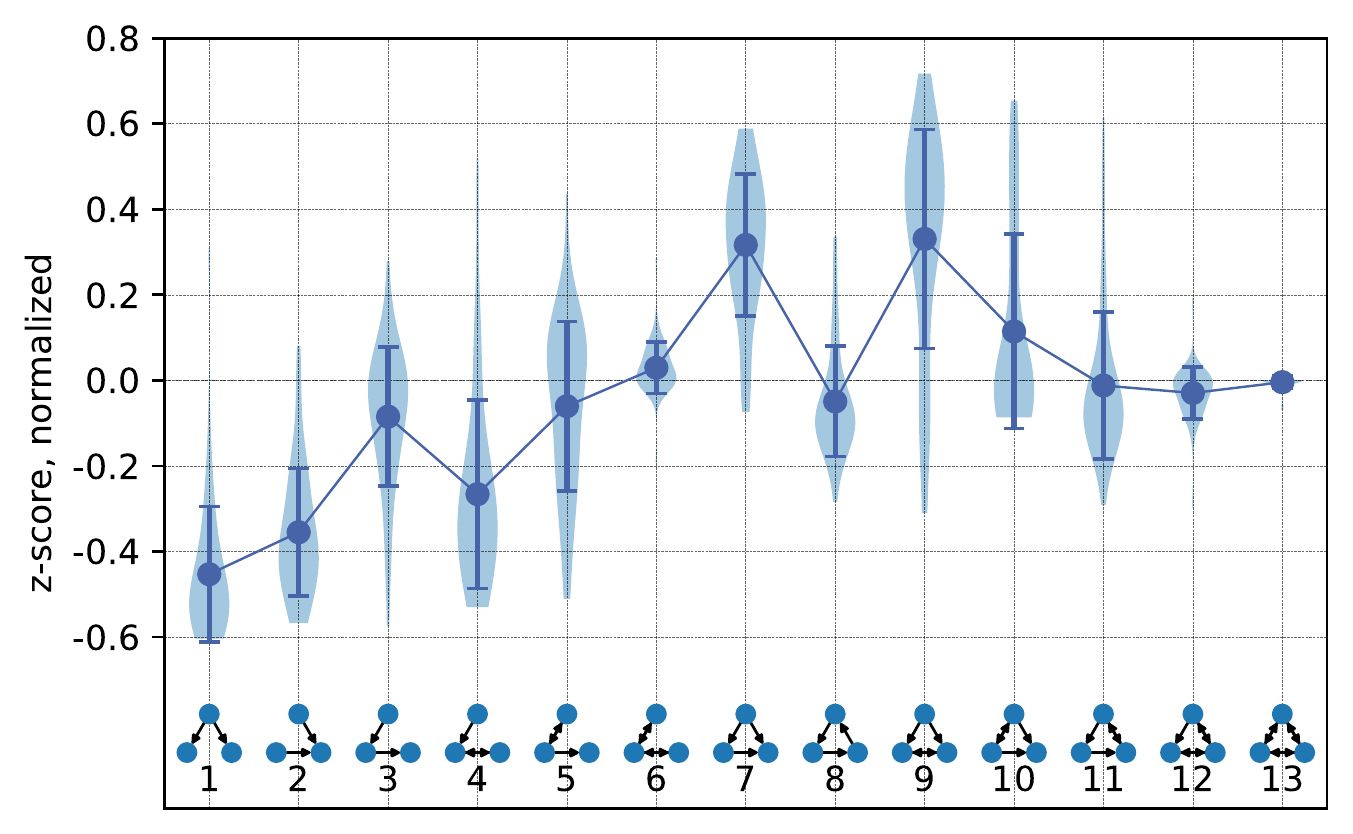}\\
	$M \in [16, 22], \robustness=0.95$
    \end{minipage}
    \begin{minipage}[t]{.45\linewidth}
        \centering
        \includegraphics[width=\columnwidth]{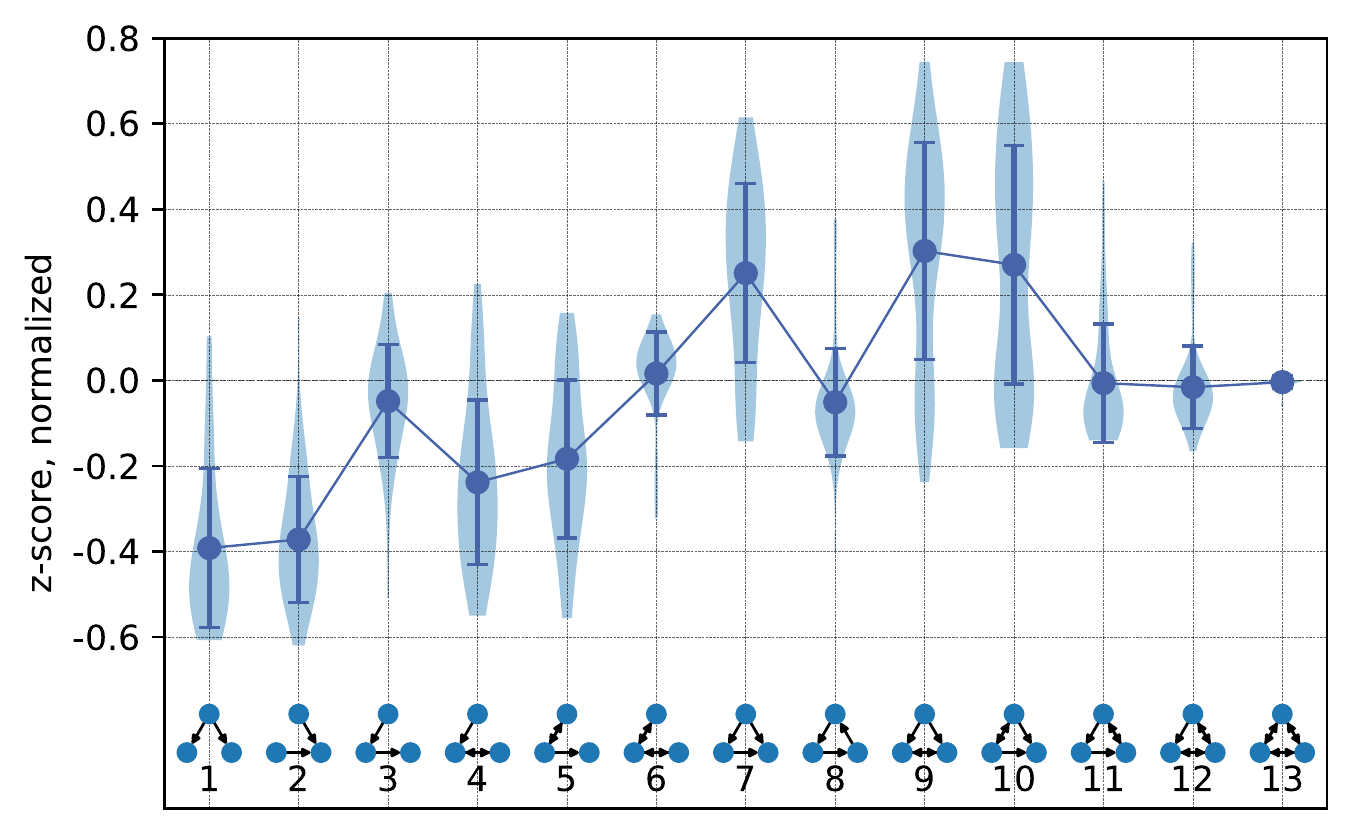}\\
	$M \in [19, 25], \robustness=0.96$
    \end{minipage}
    \caption{Motif patterns of robust networks in $\networkcost$, $\robustness$ optimization with $N=20$. Similarly to Fig.~\ref{fig:N=20-zscores-vulnerable} we take 50 networks with high $\robustness$ from the Pareto fronts.}
    \label{fig:N=20-zscores-robust}
\end{figure*}

Figures~\ref{fig:N=20-zscores-vulnerable} and \ref{fig:N=20-zscores-robust} show motif patterns of low and high robustness networks from 50 different setups. The analysis of different $M$ boundaries shows that the range between $(16, 22)$ and $(19, 25)$ yields the clearest signal for both vulnerable and robust networks. The vulnerable networks have a pattern similar to the superfamily associated with words sequences in languages from~\cite{Milo2004} or the node-robust networks in~\cite{Kaluza2007}. Simpler subgraphs in such networks are overrepresented, while the more complex structures that contribute more to the robustness are underrepresented. The main difference with the languages superfamily in the vulnerable networks is the role of the feedforward and feedback loops. The lack of feedforward loops is much less significant, while the underrepresentation of the feedback loops is the feature that distinguishes the vulnerable supply networks from the null model. As can be seen in Fig.~\ref{fig:N=20-zscores-vulnerable}, the signal of vulnerable networks disappears fast with the growth of $M$, as the low robustness becomes hard to achieve with more edges.

The robust networks have an over- and underrepresentation pattern of three-node subgraphs similar to the second superfamily from~\cite{Milo2004}. This superfamily is associated with biological networks (signaling network of living organisms, gene regulatory networks, neuronal networks). Important features of this motif signature are the neutrality of z-scores for subgraphs 3, 6, 8, the importance of the feedforward loop and its bi-directional version (7, 9). This pattern is also present across a wider range of $M$ compared to the signature of vulnerable networks. The signal, however, becomes less informative at the higher $M$ values, as installing robustness becomes easier with the help of any of the subgraphs. This appearance and decay of both high and low robustness signals can be observed in Fig.~\ref{fig:N=20-zsc-r}, where the pattern strength is plotted as the function of robustness $\robustness$. Around $M \in [16,22]$ and $[19,25]$ the signal is the strongest, showing the biggest difference between high and low robustness networks.

Similar experiments on smaller ($N=10$, Fig.~\ref{fig:N=10-zscores}) and larger ($N=30$, Fig.~\ref{fig:N=30-zscores}) networks show the same motif signatures for both robust and vulnerable networks. The peak strength of the pattern signal appears when $M \approx N$, indicating that for the investigated setup the signal strength depends on the network's average degree, rather than on the connectivity. 

\begin{figure}[ht]
    \centering
    \includegraphics[width=.6\linewidth]{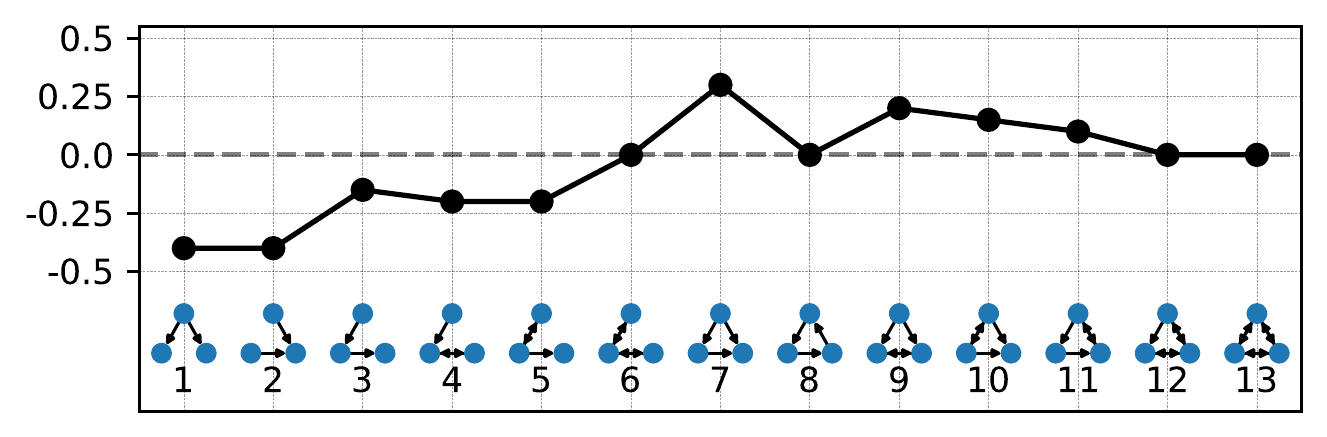}
    \caption{The target network motif pattern, reproduced from~\cite{Milo2004}.}
    \label{fig:target-z}
\end{figure}

In order to further investigate the similarity of the motif signature obtained here with the corresponding superfamily from~\cite{Milo2004}, we compute the Pearson correlation coefficient of motif signature in our model with the representation of the superfamily signature shown in Fig.~\ref{fig:target-z}. We denote this correlation coefficient the signature strength $\zsc$ and the superfamily signature as \textit{target z-score vector}. This quantity will be analyzed in detail in the following section. The resulting network parameter $\zsc$ shows how close the motif pattern of the network is to the target pattern. The value of $\zsc$ can vary between $-1$ and $1$.

\subsection{Motifs as a way to generate robust networks}
In the first part of our investigation, we have seen how the robust networks tend to have a rather precise set of local topological features. In other words, the majority of the robust networks lie in the set of networks with a particular motif signature which, in fact, has already been associated with robustness in previous studies, in the context of regulatory systems~\cite{Milo2004} and layered flow systems~\cite{Kaluza2007}. As a next step we will now address the opposite direction of this statistical association between robustness and motif signatures: If we generate a network with the required pattern, will it necessarily be more robust than the others? Or will such a network have the same high level of robustness but at a lower network cost? In the existing literature, an approach of installing robustness via generating a certain motif pattern is not well investigated, perhaps related to the computational complexity of computing network motifs, which is $O(M^4)$ ($O(N^3)$ to count subgraphs times O($M$) to get good statistics for z-scores, and having in mind that $N \approx M$ for the robustness signal), as compared to the $O(M^2)$ for the robustness ($O(M)$ is the complexity of computing the $\demandsatisfaction$ times $M$ to compute demand satisfaction $\demandsatisfaction$ without every edge). This is a relevant question, as in principle it could be that the set of networks with the pattern is bigger than the set of the robust networks and it is possible to generate a vulnerable network with the given pattern.  

In order to address this question, we perform a full optimization for the given motif pattern. Furthermore, we suggest a heuristic that indirectly installs the pattern, but does that with a smaller computational complexity.

\subsubsection*{Full motif pattern optimization}
Similar to the robustness optimization, we solve an optimization problem for 50 different setups with the random spatial distribution of nodes. The objective functions in this case are the network cost $\networkcost$ and the signature strength $\zsc$, i.e. the correlation of the network's z-score vector with the target vector given in Fig.~\ref{fig:target-z}. Due to the time complexity of optimizing $\zsc$, this experiment has been performed for a single set of edge boundaries $(M_{min}, M_{max}) = (16, 22)$ and $N=20$, i.e. the edge range for which the low and the high robustness signals were the strongest. This optimization had one additional constraint, namely that only edges not longer than a certain length $l$ were allowed to form networks. The allowed length $l$ is defined as 120\% of the length that makes the network connected. Our numerical simulations have shown that this length constraint greatly enhances the association of the motif signature with robustness. Regarding the impact of edge length, and thus spatial locality of the network, on the motif patterns one can refer to~\cite{Artzy-Randrup2004}.

\begin{figure*}[h]
    \centering
    \begin{minipage}[t]{.5\linewidth}
        \centering
        \includegraphics[width=\columnwidth]{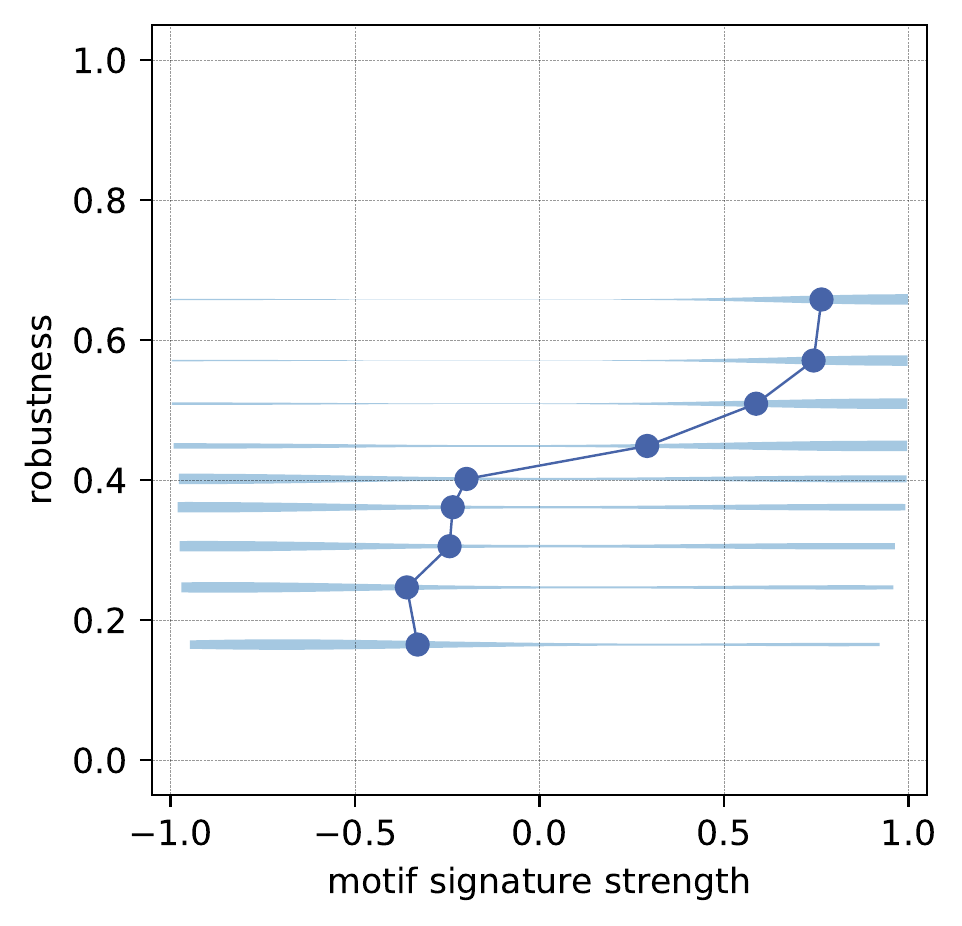}
    \end{minipage}
    \captionsetup{width=0.90\columnwidth}
    \caption{Distribution of $\robustness$ values as a function of $\zsc$ for the results of $(\networkcost, \zsc)$ optimization. Sets of networks that correspond to given robustness are plotted as violin plots to demonstrate their distributions, while the solid line shows the behavior of their means.}
    \label{fig:opt_zsc-r-vs-zsc}
\end{figure*}

From the Pareto front of this optimization (Fig.~\ref{fig:opt_zsc-pareto}) it can be seen that, although the generated networks have reached both extremely high and low $\zsc$ values, the robustness range is not entirely covered, as opposed to the $\networkcost$, $\robustness$ optimization (Fig.~\ref{fig:N=20-nc-r}). Even more striking, networks with the opposite motif pattern (negative $\zsc$ values in Fig.~\ref{fig:opt_zsc-pareto}) reach robustness values that are comparable to those with target pattern. This observation indicates that not all networks with the given motif pattern will necessarily be robust. On the other hand, looking at the robustness as the function of motif patterns (Fig.~\ref{fig:opt_zsc-r-vs-zsc}), it can be seen that high robustness networks have almost exclusively the target motif pattern.

\subsubsection*{Heuristic motif pattern enhancement}
The approach with direct motif pattern optimization has proven to be both computationally complex and inefficient in generating robust networks. Here we try to address these problems by testing a heuristic that follows simple rules based on motif patterns. In this approach, we consider a more practical setup, with a randomly given network that has $\demandsatisfaction=1$. The goal is to insert one edge into the network such that the insertion brings the existing motif signature closer towards the target signature, increasing the robustness in the process.

\begin{figure*}[h]
    \centering
    \begin{minipage}[t]{.3\linewidth}
        \centering
        \includegraphics[height=\columnwidth]{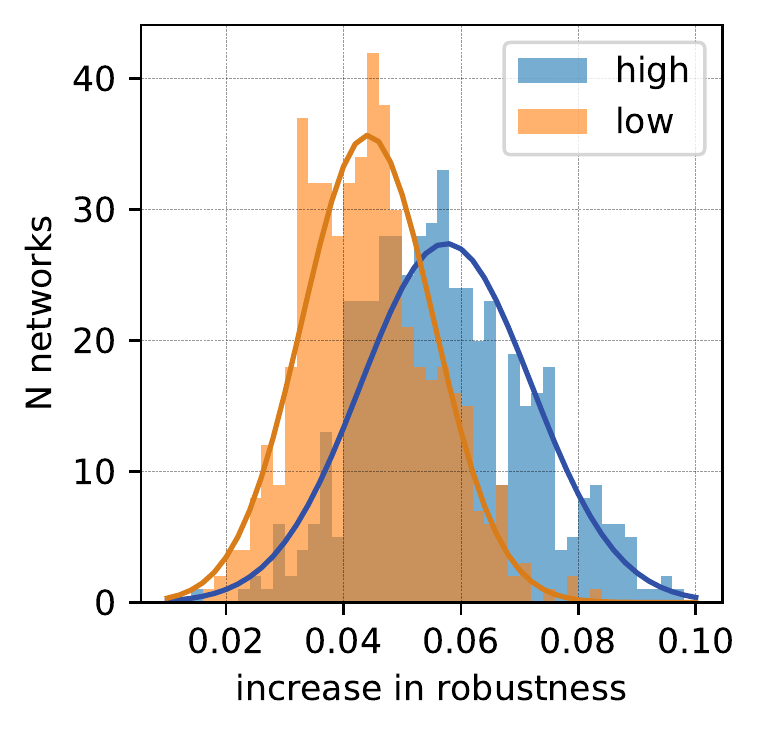}
        \subcaption{FFL counts}
        \label{fig:heur_zsc-means-c07}
    \end{minipage}
    \begin{minipage}[t]{.3\linewidth}
        \centering
        \includegraphics[height=\columnwidth]{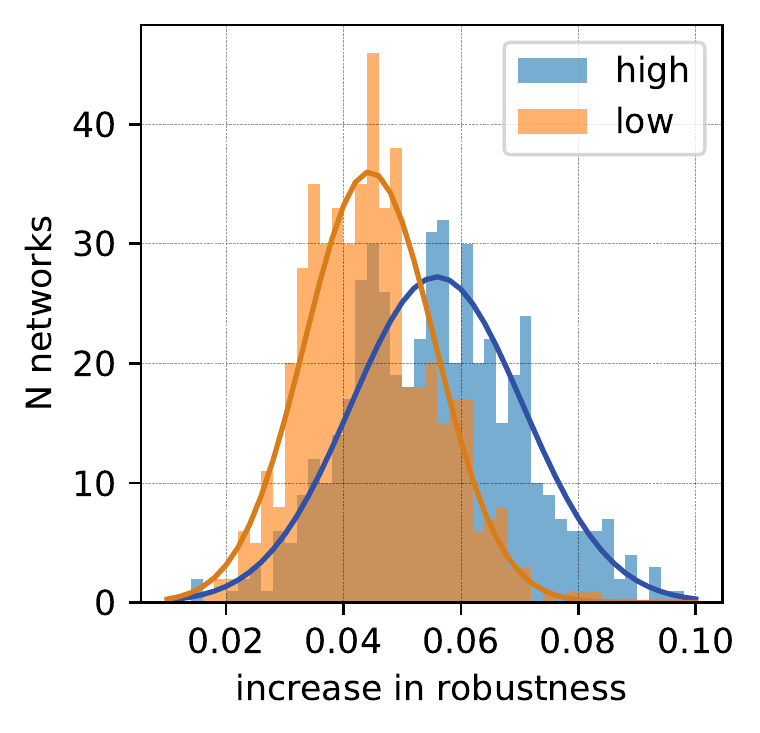}
        \subcaption{FFL z-scores}
        \label{fig:heur_zsc-means-z07}
    \end{minipage}
    \begin{minipage}[t]{.3\linewidth}
        \centering
        \includegraphics[height=\columnwidth]{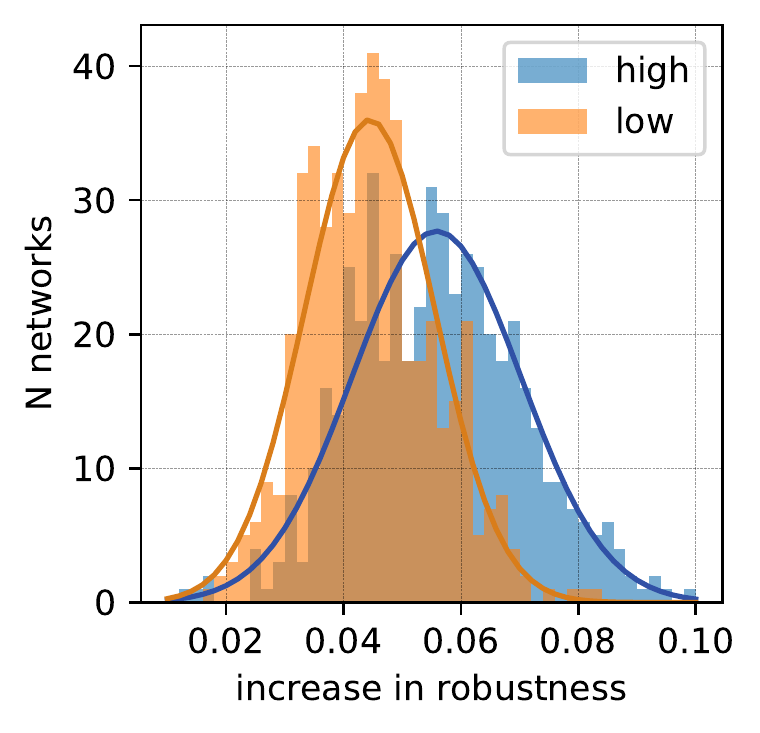}
        \subcaption{$\zsc$}
        \label{fig:heur_zsc-means-zsc}
    \end{minipage}
    \captionsetup{width=0.90\columnwidth}
    \caption{Histograms of the average increase in robustness for high and low changes in one topological metric. We take 500 random networks that have full demand satisfaction and some robustness. For each combination of a network and an edge that is not in the network, we compute $r$, motif signature, and three-node subgraph counts. Then, for each network, we separate combinations into two groups: the ones that yield the highest increase in a simple metric (counts of feedforward loop (a), z-score of feedforward loop (b), and $\zsc$ (c)) -- denoted as high, and the remaining combinations -- denoted as low. After this we compare the average increase in $r$ in these two groups.}
    \label{fig:heur_zsc-means}
\end{figure*}

We investigate this problem by taking 500 different random networks with $N=20, M \in [16, 22]$ and inserting every edge that did not exist in the original network. The extended networks are then compared with the basic one by three parameters: count of feedforward loops ($c07$), z-score of the feedforward loop ($z07$), and overall correlation with the target z-score pattern ($\zsc$). After that, we split all the possible networks into two groups: \textbf{high} -- those that give the maximum increase in $c07$, $z07$, and $\zsc$ and \textbf{low} -- the remaining networks. In the case of $c07$, the high group is composed of networks with positive change in the 7th subgraph counts. In the case of $z07$ and $\zsc$, the top 10\% of the networks are taken (see more details in the Supplementary Information, Fig.~\ref{fig:heur_zsc-changes}). Then we compare the increase in robustness in these two groups taking their mean values. This process is then repeated 500 times for different base networks and the histogram of high and low means is finally plotted (Fig.~\ref{fig:heur_zsc-means}). 

The resulting figure tells us that all three parameters work equally well for edge insertion. Selecting the edges with the highest values of $c07$, $z07$, or $\zsc$ leads to a higher increase in robustness. To our knowledge, these two investigation steps -- the full motif pattern optimization and the heuristic motif pattern enhancement -- are the first examples indicating that indeed this motif signature \textit{implies} robustness. However, in both numerical experiments, the observed effect is weak. Furthermore, even the heuristic is still more computationally complex than the direct robustness computations, even using the simplest $c07$ approach. These observations show that the results, though of relevance for the theoretical understanding of supply network robustness, will most likely not be of immediate practical relevance. One possible application of such a heuristic might be in the case when computing or defining robustness is hard.

\subsection{Application to industrial data}
\begin{figure*}[h]
    \centering
    \begin{minipage}[t]{.58\linewidth}
		\centering
		\includegraphics[width=\columnwidth]{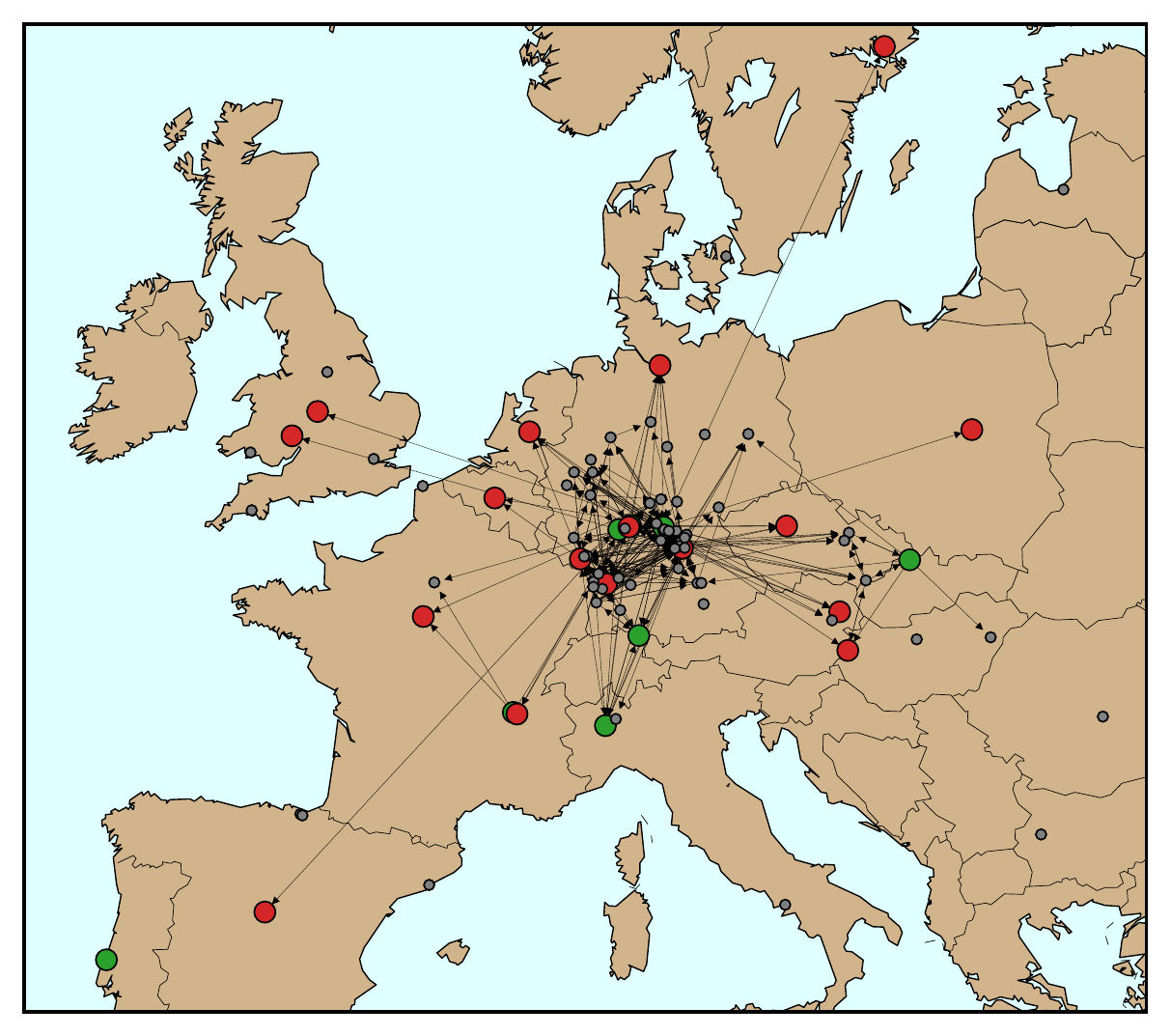}
		\subcaption{}
		\label{fig:real-nw}
	\end{minipage}
	\begin{minipage}[t]{.38\linewidth}
		\centering
		\includegraphics[width=\columnwidth]{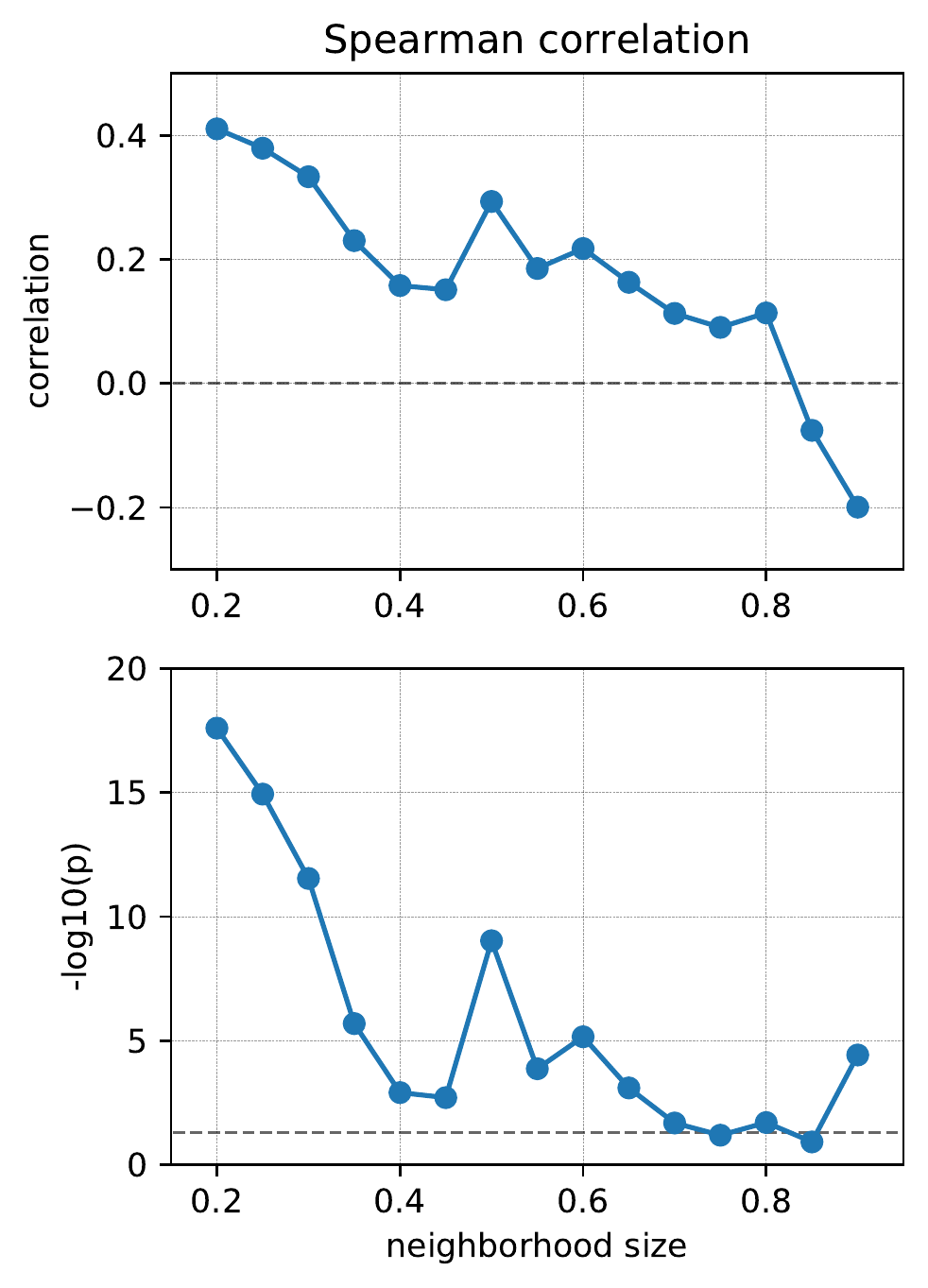}
		\subcaption{}
		\label{fig:real-zsc-r-spearman}
	\end{minipage}
	\captionsetup{width=0.90\columnwidth}
	\caption{Application of the model to industrial data. Figure (a) shows a subnetwork of the full European supply network of an automotive company. The subnetwork is generated by selecting the routes used for transporting one product category and taking a neigborhood of these routes with size parameter $t=0.4$. In Fig.~(b) all product subnetworks are analysed together using Spearman correlation (upper panel) and the corresponding p-value (represented as $-\mbox{log}(p)$; lower panel) of their $\robustness$ and $\zsc$ values for different neighborhood sizes $t$. The dashed lines indicate zero correlation (upper panel) and $-\mbox{log}(0.05)$ (lower panel), respectively.}
    \label{fig:real-joined}
\end{figure*}

Next, we apply the concept of the model to a real-world supply network. As an example, we take the transportation network of a global automotive supplier and create a map of all existing routes among all European facilities. These facilities both produce and demand products and the whole network is an overlay of a large product portfolio. Therefore we split the network into different subnetworks that operate for a single product category only. For a single product category, all deliveries happen from the supplier to the demander directly and without additional edges that provide system robustness. In reality, as the whole transportation network consists of multiple products and deliveries are combined together, the actual network for a single product includes additional routes. To model this behavior for each product we consider the neighborhood area of the original direct routes. This is achieved by taking all possible paths from supplier nodes that are no longer than the local neighborhood size parameter $t$ times the length of the direct link, as well as considering not more than 3 edges in a single path. After this it is possible to apply the model and compute most of the necessary network parameters $\demandsatisfaction$, $\robustness$, $\zsc$. As a result, we have 627 product subnetworks, each composed of producing and demanding nodes (see an example of a product subnetwork in Fig.~\ref{fig:real-nw}).

A Spearman correlation analysis of the two quantities in Fig.~\ref{fig:real-zsc-r-spearman}, robustness and motif signature strength, shows a strong $(> 0.3)$ and significant $(p-value << 0.05)$ for small neighborhood sizes. This correlation becomes weaker and then disappears with increasing neighborhood size. Analysis of the Pearson correlation coefficient confirms these observations (see Fig.~\ref{fig:real-zsc-r-pearson}) but is less reliable due to the non-Gaussian distributions of the quantities investigated. As we are going to the region of bigger neighborhood sizes, product sub-networks include more and more edges and become closer to one another, making any analysis meaningless. 

Overall, this experiment is encouraging as it shows that it is possible to view the real-world networks from the prism of the model. However, to gain any meaningful insight on an operational level
 about the explored systems the model needs to be further developed, simulating setups closer to the practical networks.

\section{Conclusion}
Here we have presented a minimal model of supply networks. Although the model is based on one simple mechanism of matching the demand and supply, it proves to be powerful in describing the concepts of robustness and efficiency in supply networks. The optimal networks generated in this framework show structural patterns that are also typical to biological systems. This finding unites the nature of two network worlds -- those found in natural systems that have been developed under evolutionary processes and industrial systems that are artificially created with the main goal of being cost-efficient. Having this evidence of similarity, it is possible to explore both systems from the perspective of another and potentially transfer the knowledge between them.

One of the questions addressed in our investigation is whether one can use motif patterns as a building recipe for robust networks. This approach turns out to be much more complex on the computational side, while also showing only a weak benefit in comparison to a random pattern. This numerical observation indicates that the family of networks with the given motif pattern is wider and includes not only the robust networks. An important finding, however, is that requiring a spatial locality of the network edges forces robust networks to adhere to the given motif signature. The discussed motif pattern thus should not be associated with robustness in isolation but should be augmented by some additional factors such as spatial aggregation. 

Finally, while the suggested supply network model is minimal, it has substantial potential for further investigations. The simplest direction is to apply different distance measures to compute network cost $\networkcost$. For example, instead of the standard Euclidean distance edges that are shorter than some threshold distance $x$ can have a length of 1, while the longer edges can have an infinite length. This will model the situation when the delivery vehicles can travel no longer than $x$ per one go. Another direction is to go beyond the single-product dimension. In the explored setups there was only one producing node which is usually not the case in reality. Also, the spatial distribution of the nodes is not uniform, especially in the case of worldwide supply networks. An even more interesting setup can have an overlay of different products, each with its own demanders and producers. In this setup, some edges will be more valuable as they can enhance the robustness more efficiently. 

With this minimal model, we hope to provide an interface between the multidisciplinary field of network science and research questions in supply network management. 

\bibliographystyle{unsrt}  
\bibliography{references}

\clearpage
\newcommand{\beginsupplement}{%
    \setcounter{table}{0}
    \renewcommand{\thetable}{S\arabic{table}}%
    \setcounter{figure}{0}
    \renewcommand{\thefigure}{S\arabic{figure}}%
}
\beginsupplement
\section*{Supplementary Information}

\begin{figure*}[ht]
    \centering
    \begin{minipage}[t]{.27\linewidth}
        \centering
        \includegraphics[width=\columnwidth]{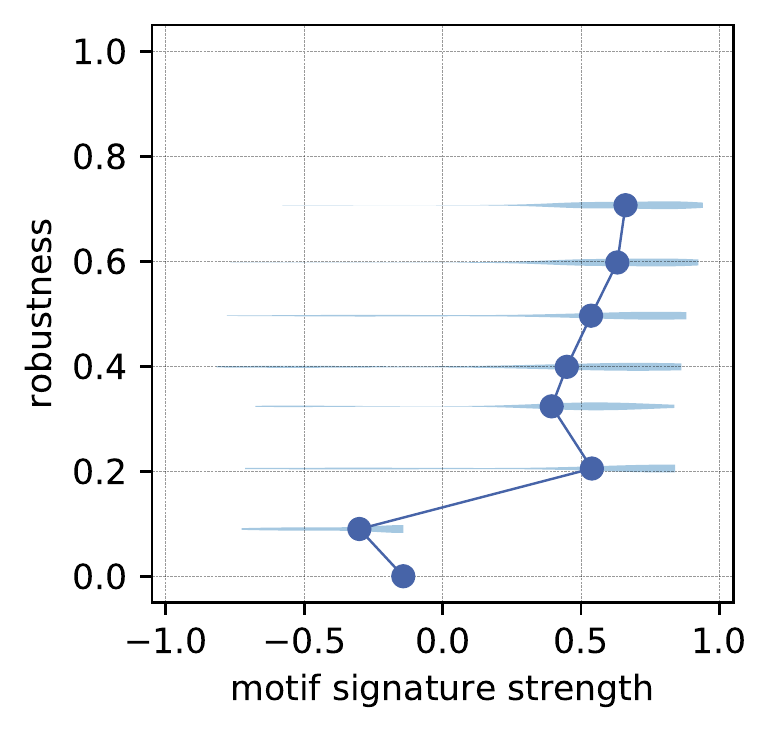}\\
	$M \in [10, 16]$
    \end{minipage}
    \begin{minipage}[t]{.27\linewidth}
        \centering
        \includegraphics[width=\columnwidth]{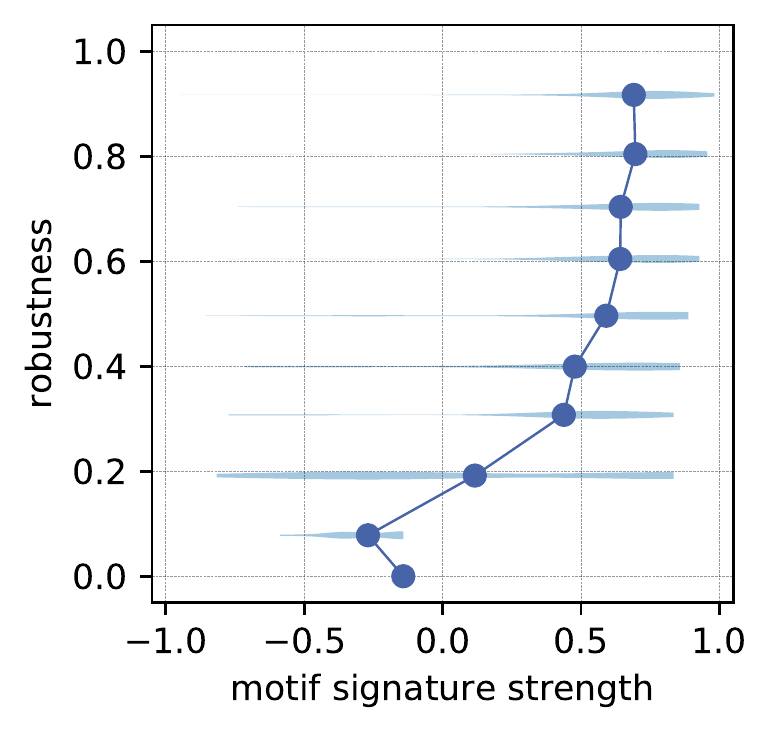}\\
	$M \in [13, 19]$
    \end{minipage}
    \begin{minipage}[t]{.27\linewidth}
        \centering
        \includegraphics[width=\columnwidth]{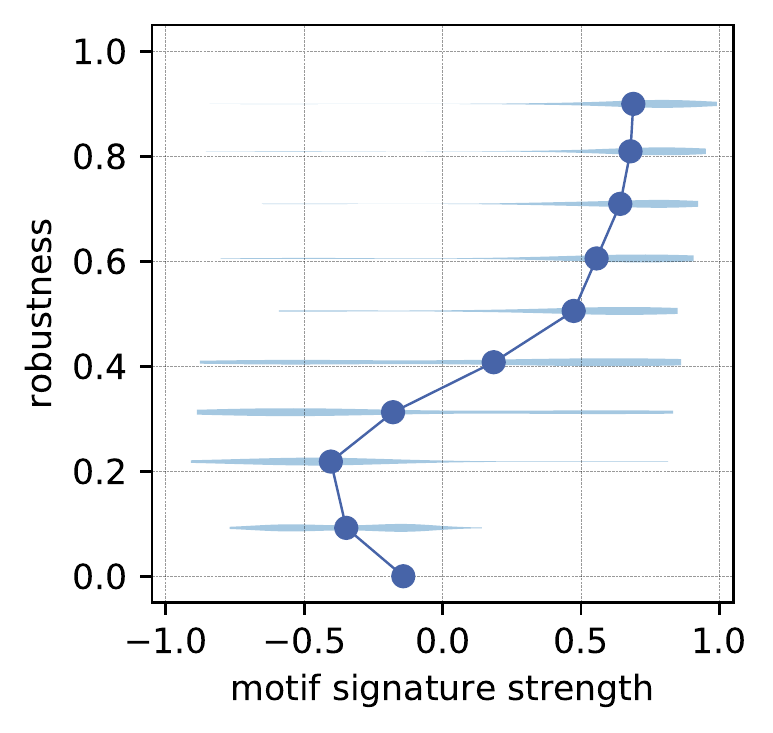}\\
	$M \in [16, 22]$
    \end{minipage} \\
    \centering
    \begin{minipage}[t]{.27\linewidth}
        \centering
        \includegraphics[width=\columnwidth]{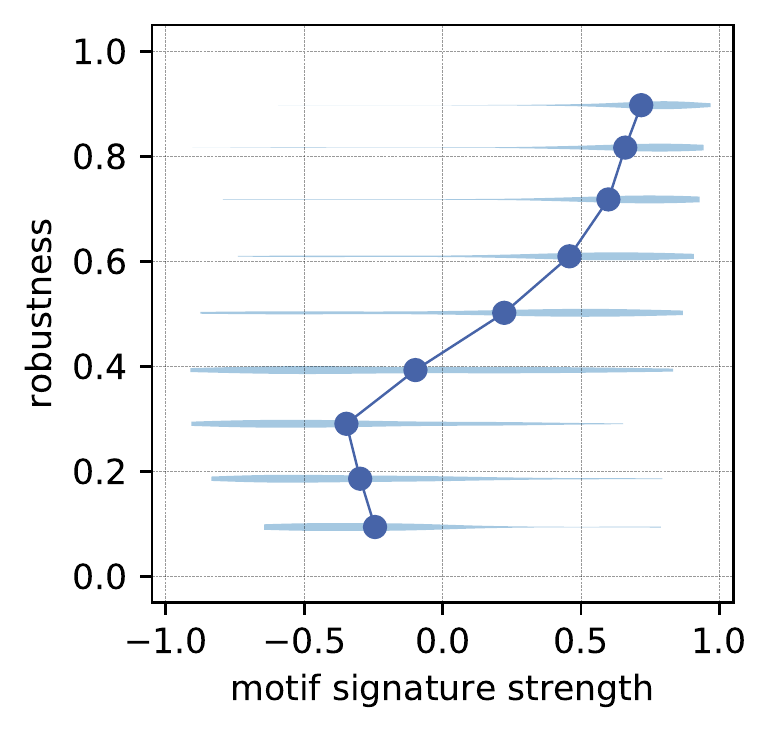}\\
	$M \in [19, 25]$
    \end{minipage}
    \begin{minipage}[t]{.27\linewidth}
        \centering
        \includegraphics[width=\columnwidth]{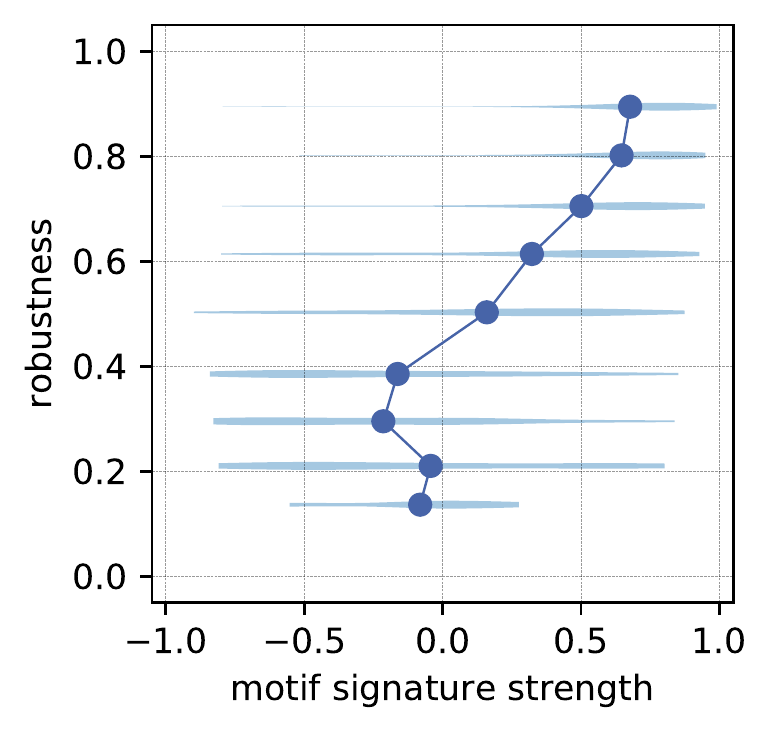}\\
	$M \in [22, 28]$
    \end{minipage}
    \begin{minipage}[t]{.27\linewidth}
        \centering
        \includegraphics[width=\columnwidth]{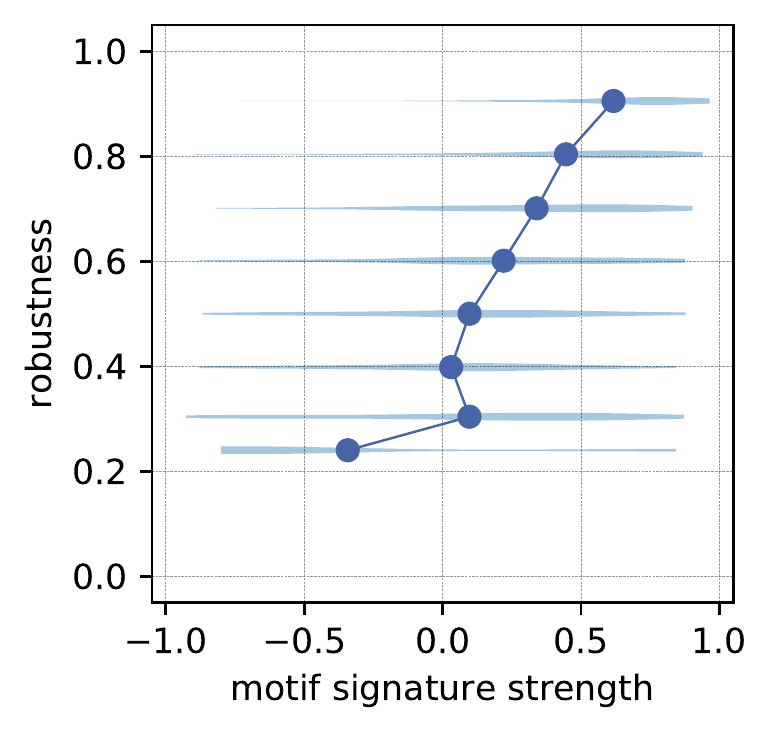}\\
	$M \in [25, 31]$
    \end{minipage} \\
    \centering
    \begin{minipage}[t]{.27\linewidth}
        \centering
        \includegraphics[width=\columnwidth]{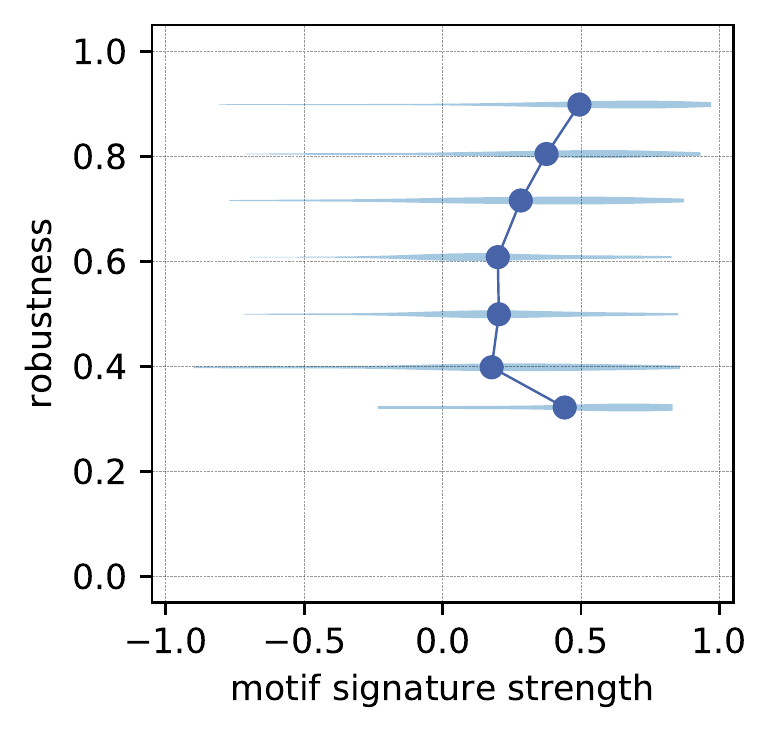}\\
	$M \in [28, 34]$
    \end{minipage}
    \begin{minipage}[t]{.27\linewidth}
        \centering
        \includegraphics[width=\columnwidth]{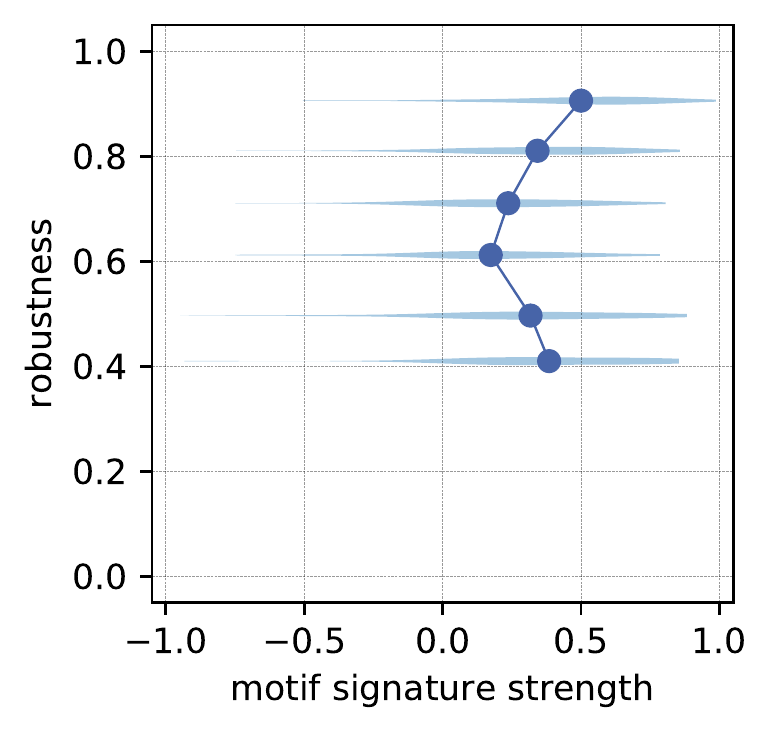}\\
	$M \in [31, 37]$
    \end{minipage}
    \begin{minipage}[t]{.27\linewidth}
        \centering
        \includegraphics[width=\columnwidth]{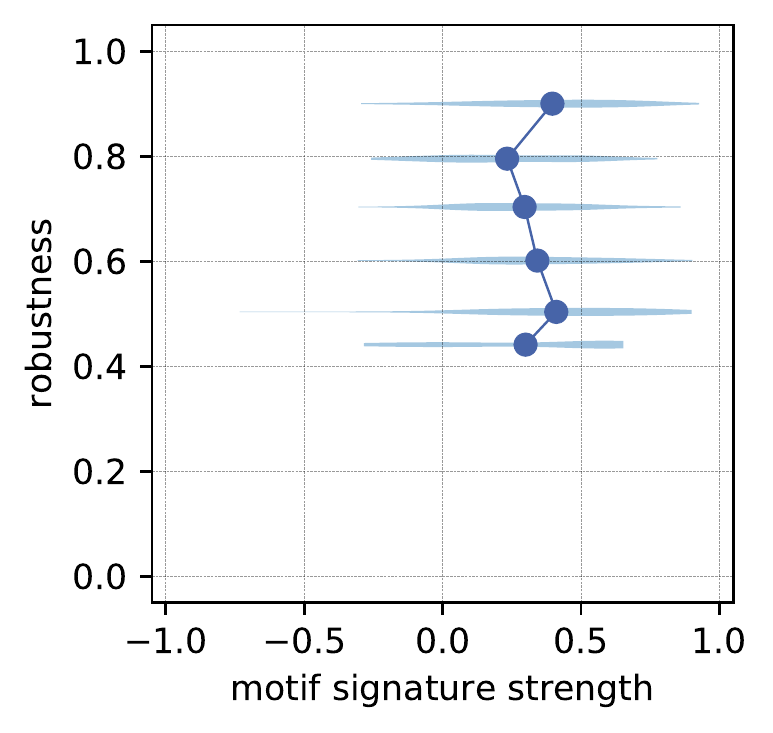}\\
	$M \in [34, 40]$
    \end{minipage} \\
    \centering
    \begin{minipage}[t]{.27\linewidth}
        \centering
        \includegraphics[width=\columnwidth]{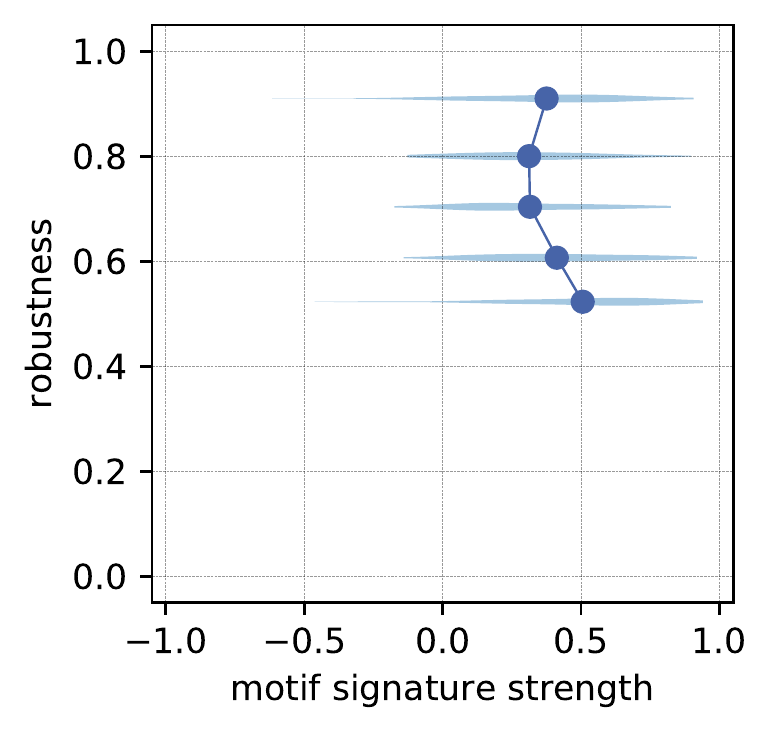}\\
	$M \in [37, 43]$
    \end{minipage}
    \begin{minipage}[t]{.27\linewidth}
        \centering
        \includegraphics[width=\columnwidth]{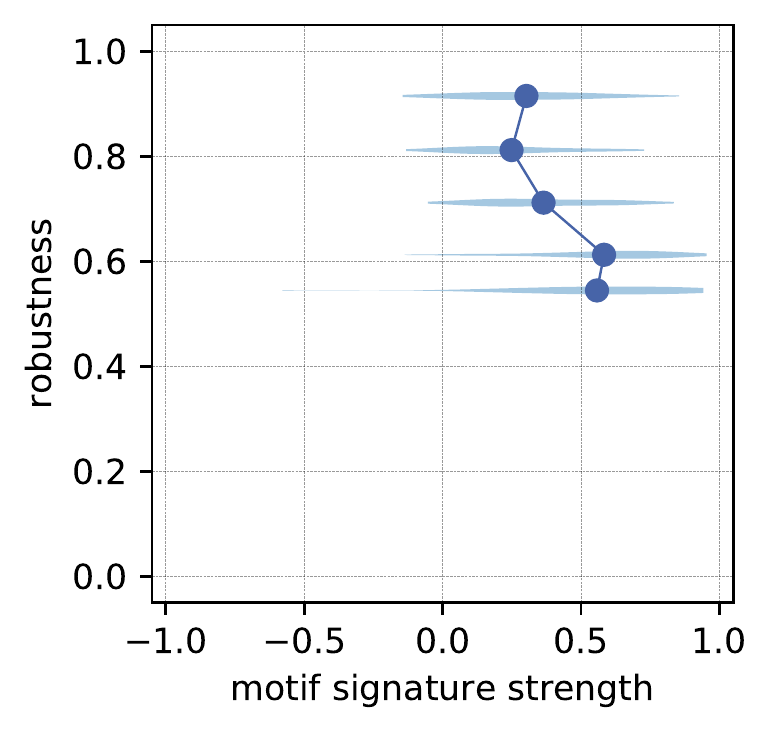}\\
	$M \in [40, 46]$
    \end{minipage}
    \caption{Distributions of $\zsc$ values for different target $\robustness$. Results from $(\networkcost, \robustness)$ optimization with $N=20$. }
    \label{fig:N=20-zsc-r}
\end{figure*}

\begin{figure*}[h]
    \centering
    \begin{minipage}[t]{.45\linewidth}
        \centering
        \includegraphics[width=\columnwidth]{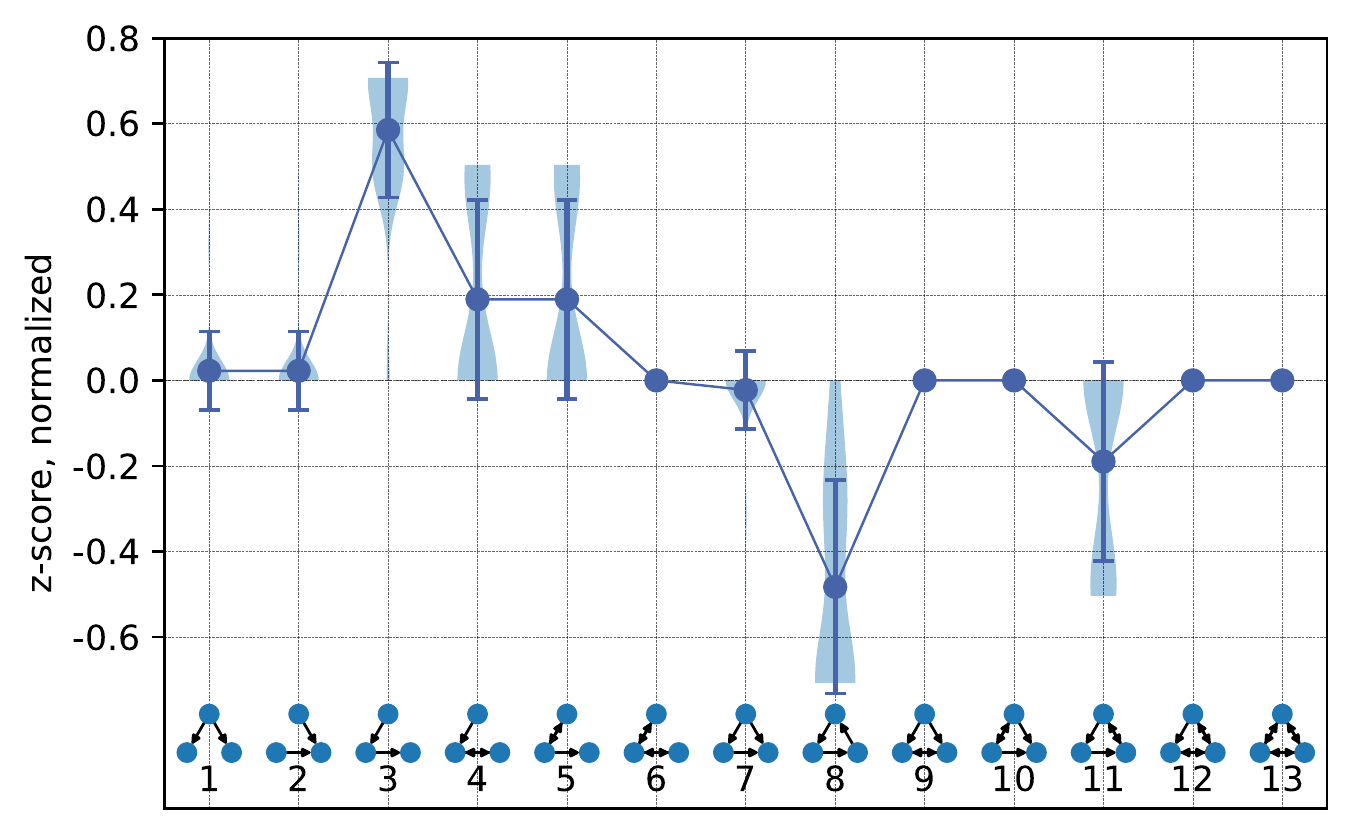}\\
	$M \in [8, 14]$, $\robustness=0.12$
    \end{minipage}
    \begin{minipage}[t]{.45\linewidth}
        \centering
        \includegraphics[width=\columnwidth]{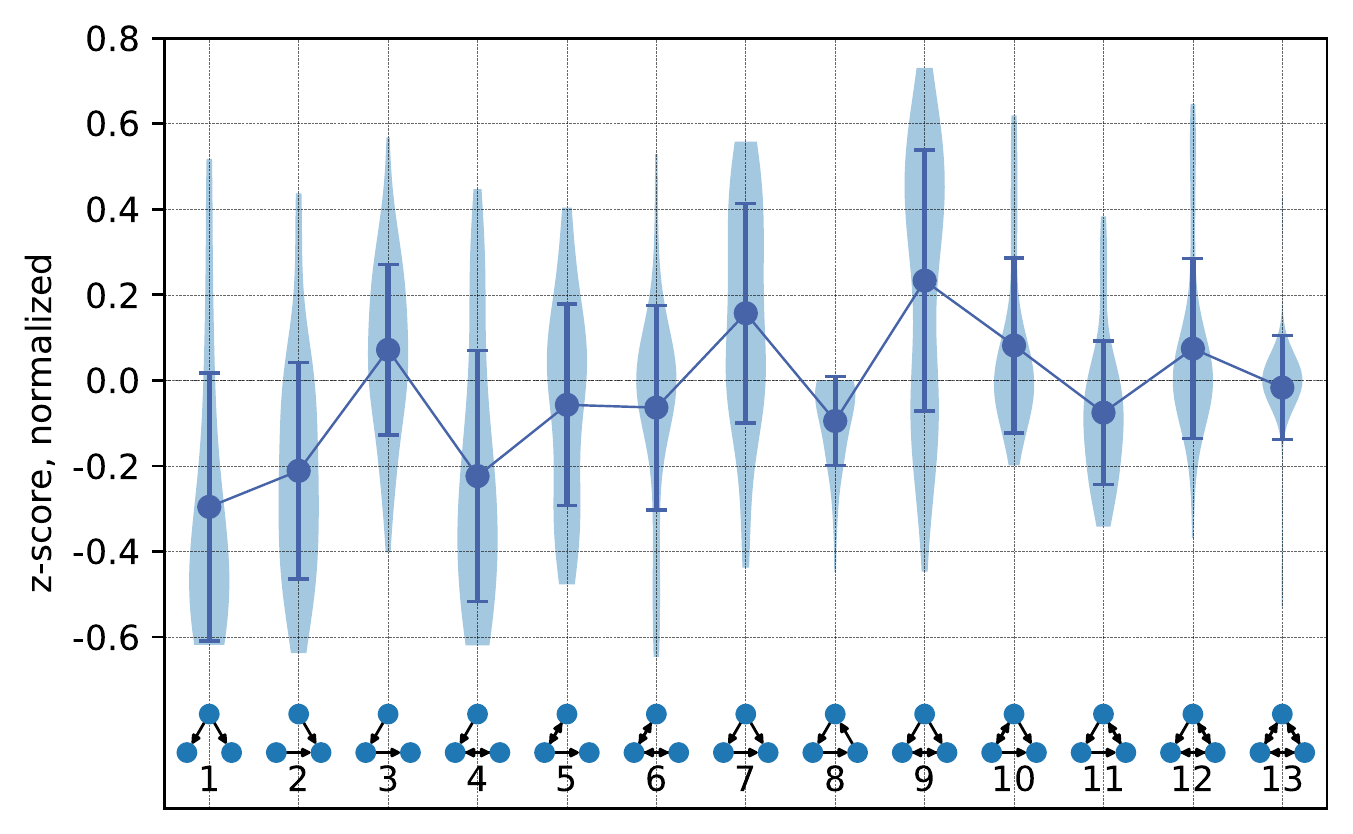}\\
	$M \in [8, 14]$, $\robustness=0.92$
    \end{minipage}
    \caption{Motif patterns in $(\networkcost, \robustness)$ optimization with $N=10$.}
    \label{fig:N=10-zscores}
\end{figure*}

\begin{figure*}[h]
    \centering
    \begin{minipage}[t]{.45\linewidth}
        \centering
        \includegraphics[width=\columnwidth]{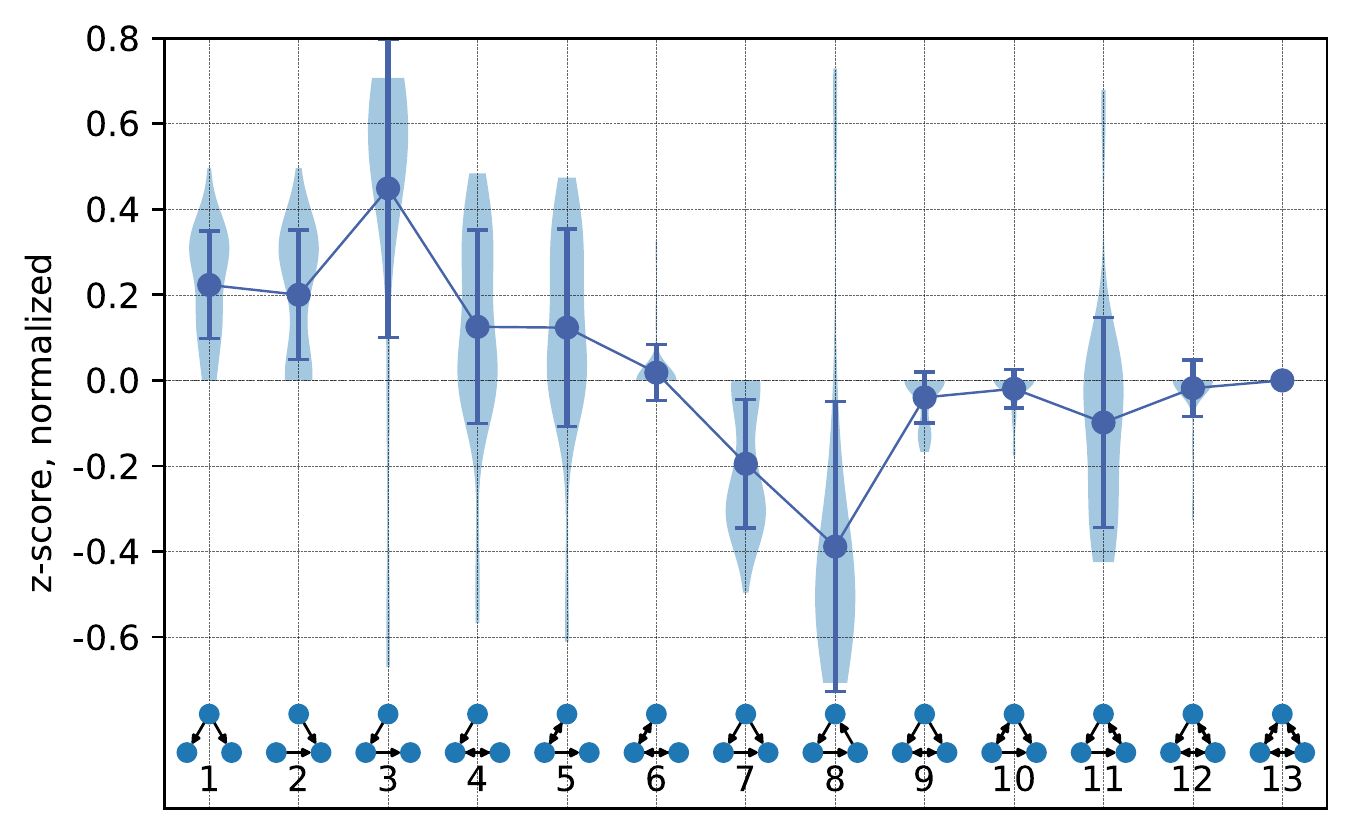}\\
	$M \in [22, 28]$, $\robustness=0.18$
    \end{minipage}
    \begin{minipage}[t]{.45\linewidth}
        \centering
        \includegraphics[width=\columnwidth]{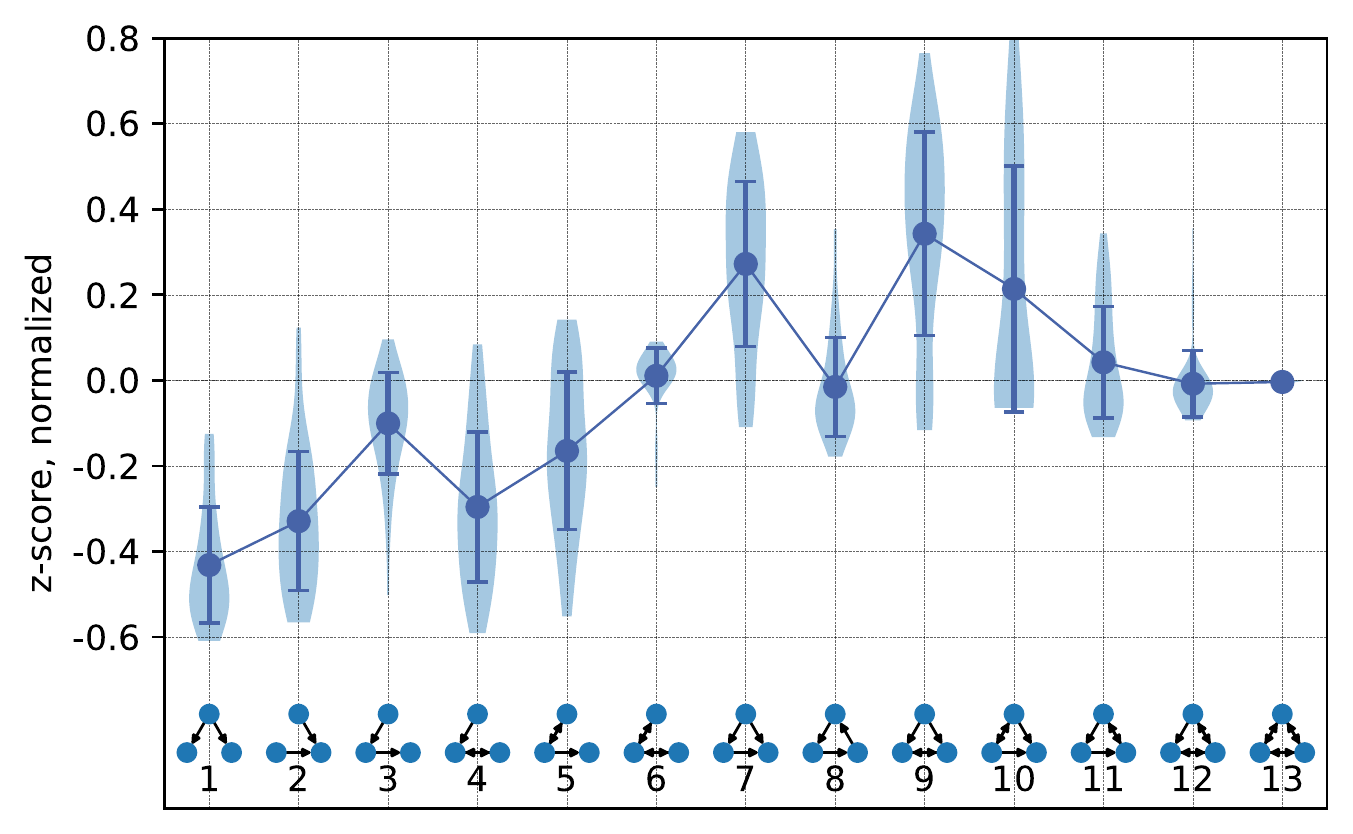}\\
	$M \in [28, 34]$, $\robustness=0.97$
    \end{minipage}
    \caption{Motif patterns in $(\networkcost, \robustness)$ optimization with $N=30$.}
    \label{fig:N=30-zscores}
\end{figure*}

\begin{figure*}[h]
    \centering
    \begin{minipage}[t]{.3\linewidth}
        \centering
        \includegraphics[height=.9\columnwidth]{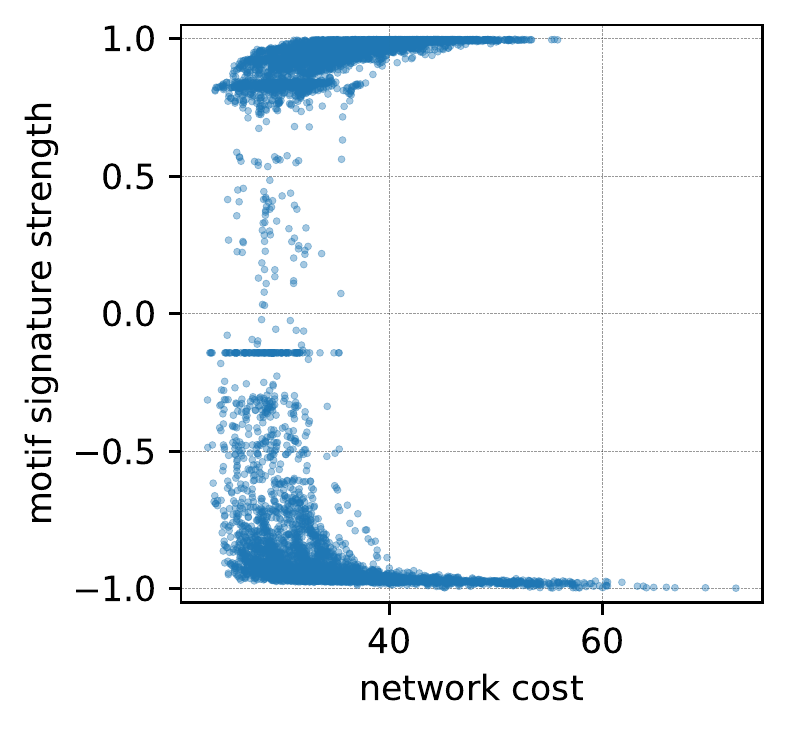}
    \end{minipage}
    \begin{minipage}[t]{.3\linewidth}
        \centering
        \includegraphics[height=.9\columnwidth]{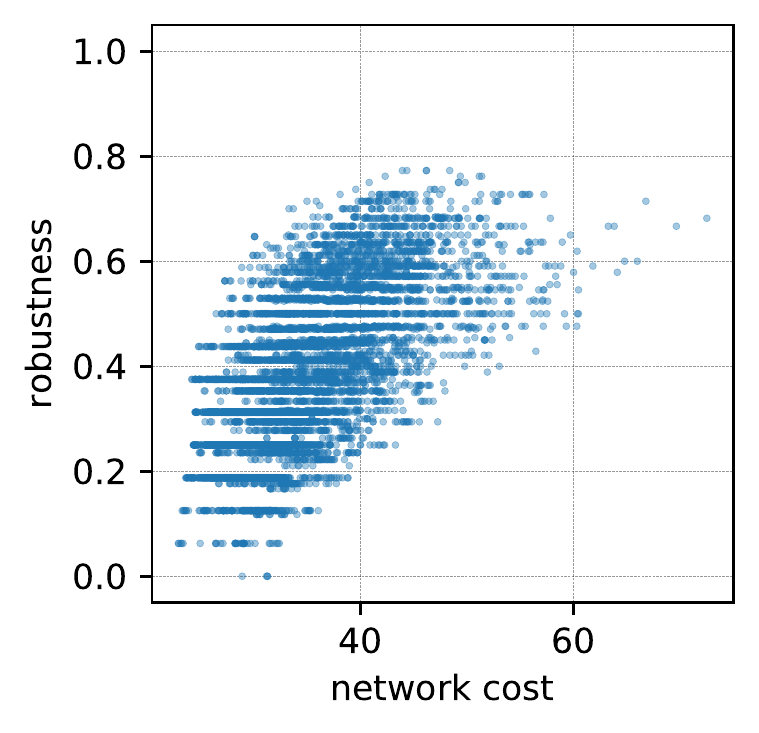}
    \end{minipage}
    \begin{minipage}[t]{.3\linewidth}
        \centering
        \includegraphics[width=\columnwidth]{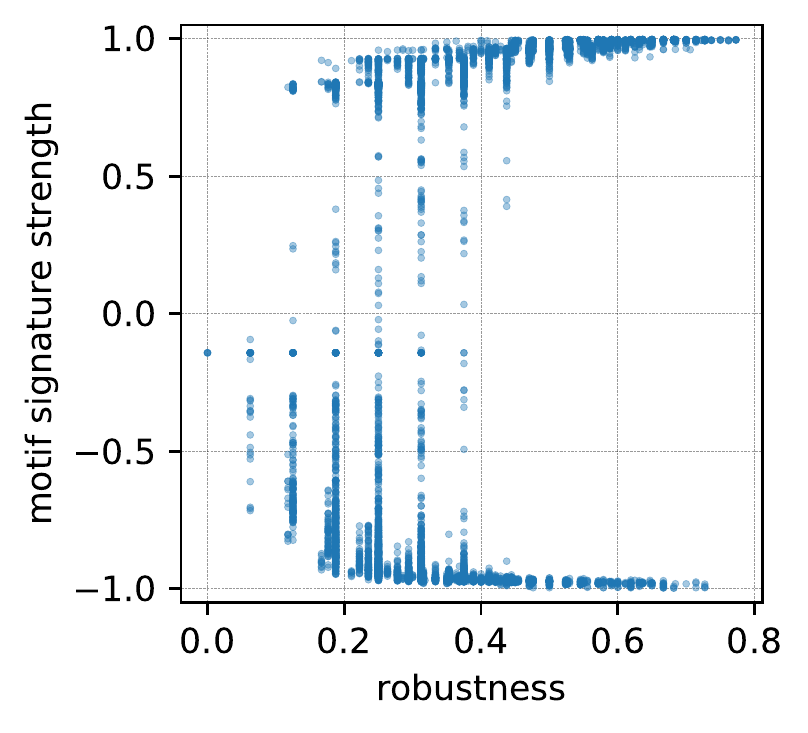}
    \end{minipage}
    \caption{Pareto fronts in $\networkcost$, $\zsc$ optimization with $N=20$.}
    \label{fig:opt_zsc-pareto}
\end{figure*}

\begin{figure*}[h]
    \centering
    \begin{minipage}[t]{.3\linewidth}
        \centering
        \includegraphics[height=\columnwidth]{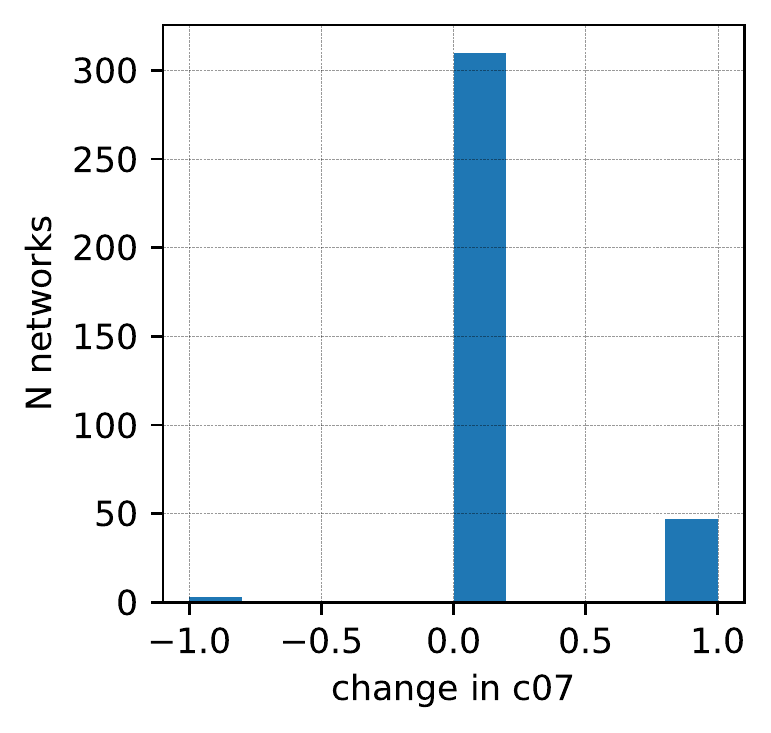}
        \subcaption{FFL counts}
        \label{fig:heur_zsc-changes-c07}
    \end{minipage}
    \begin{minipage}[t]{.3\linewidth}
        \centering
        \includegraphics[height=\columnwidth]{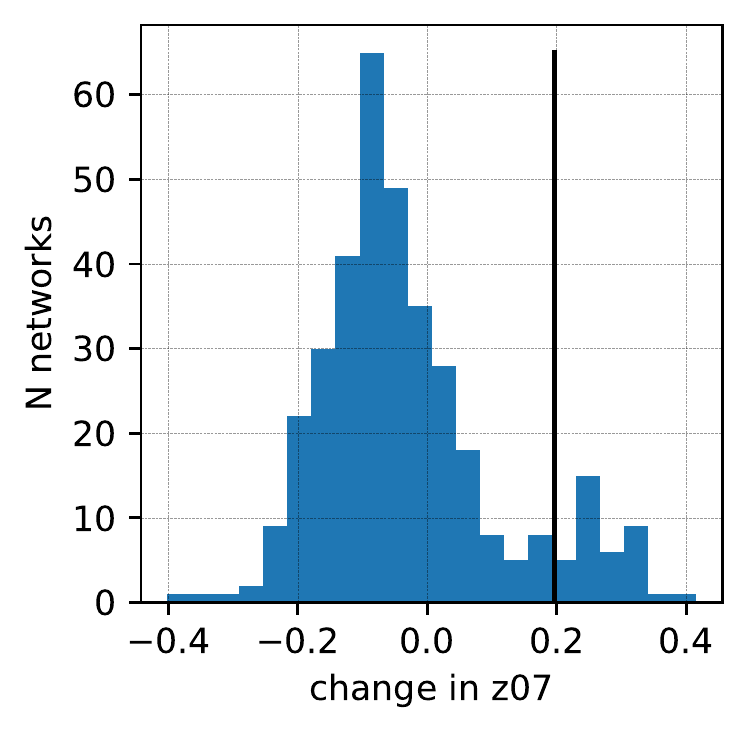}
        \subcaption{FFL z-scores}
        \label{fig:heur_zsc-changes-z07}
    \end{minipage}
    \begin{minipage}[t]{.3\linewidth}
        \centering
        \includegraphics[height=\columnwidth]{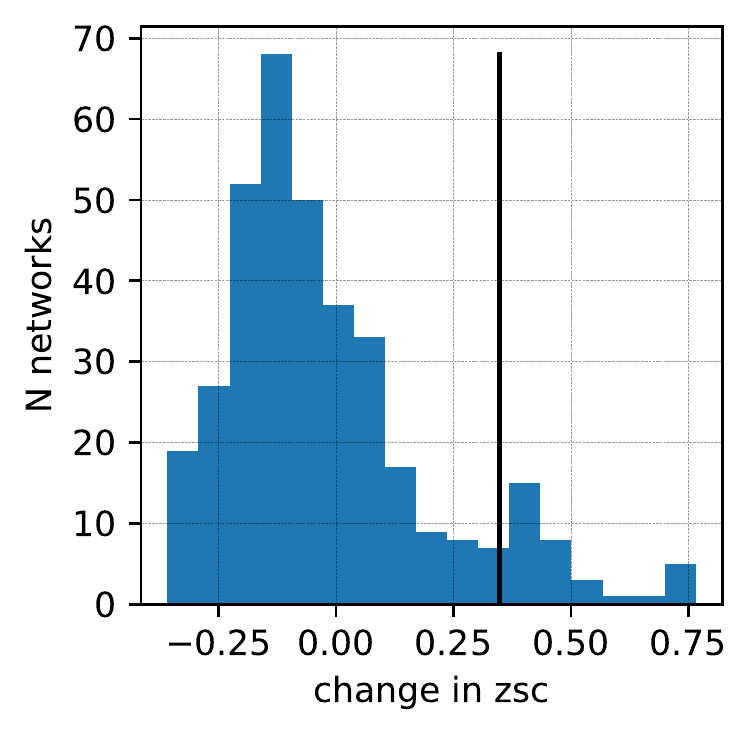}
        \subcaption{$\zsc$}
        \label{fig:heur_zsc-changes-zsc}
    \end{minipage}
    \caption{Change in $c07$, $z07$, and $\zsc$ for one base random network. Each possible edge insertion results in an extended network with new $c07$, $z07$, and $\zsc$ values. Then the histogram of the difference between these new values and the base network value is plotted. Extended networks with the highest change (above the black threshold line) are considered the best.}
    \label{fig:heur_zsc-changes}
\end{figure*}

\begin{figure*}[h]
    \centering
    \begin{minipage}[t]{.59\linewidth}
		\centering
		\includegraphics[width=\columnwidth]{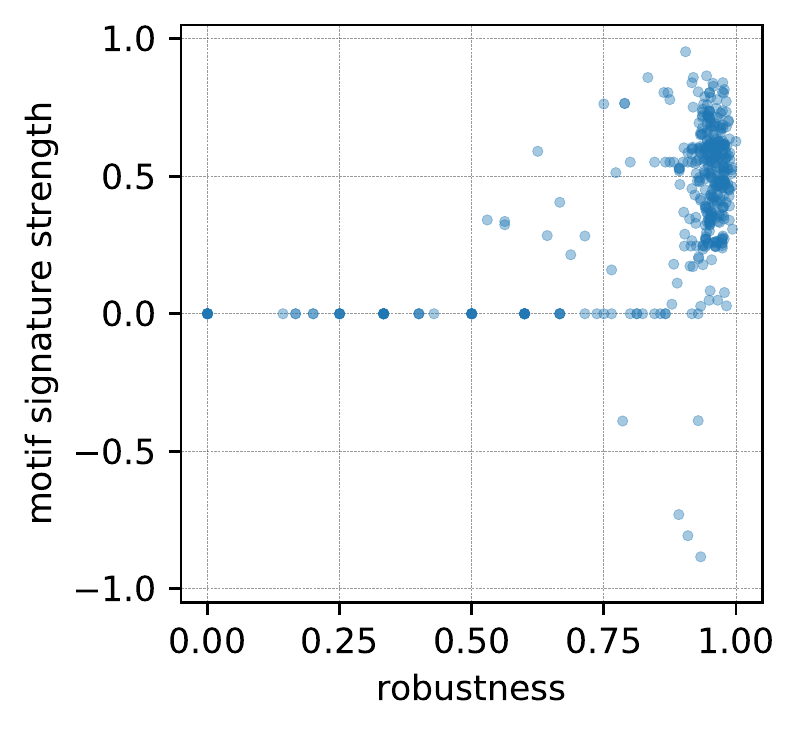}
		\captionsetup{width=0.90\columnwidth}
		\caption{Product subnetworks of the real-world supply network in the $\robustness$, $\zsc$ parameter space. Local neighborhood size parameter $t=0.2$.}
		\label{fig:real-nw-r-zsc}
    \end{minipage}
    \begin{minipage}[t]{.40\linewidth}
		\centering
		\includegraphics[width=\columnwidth]{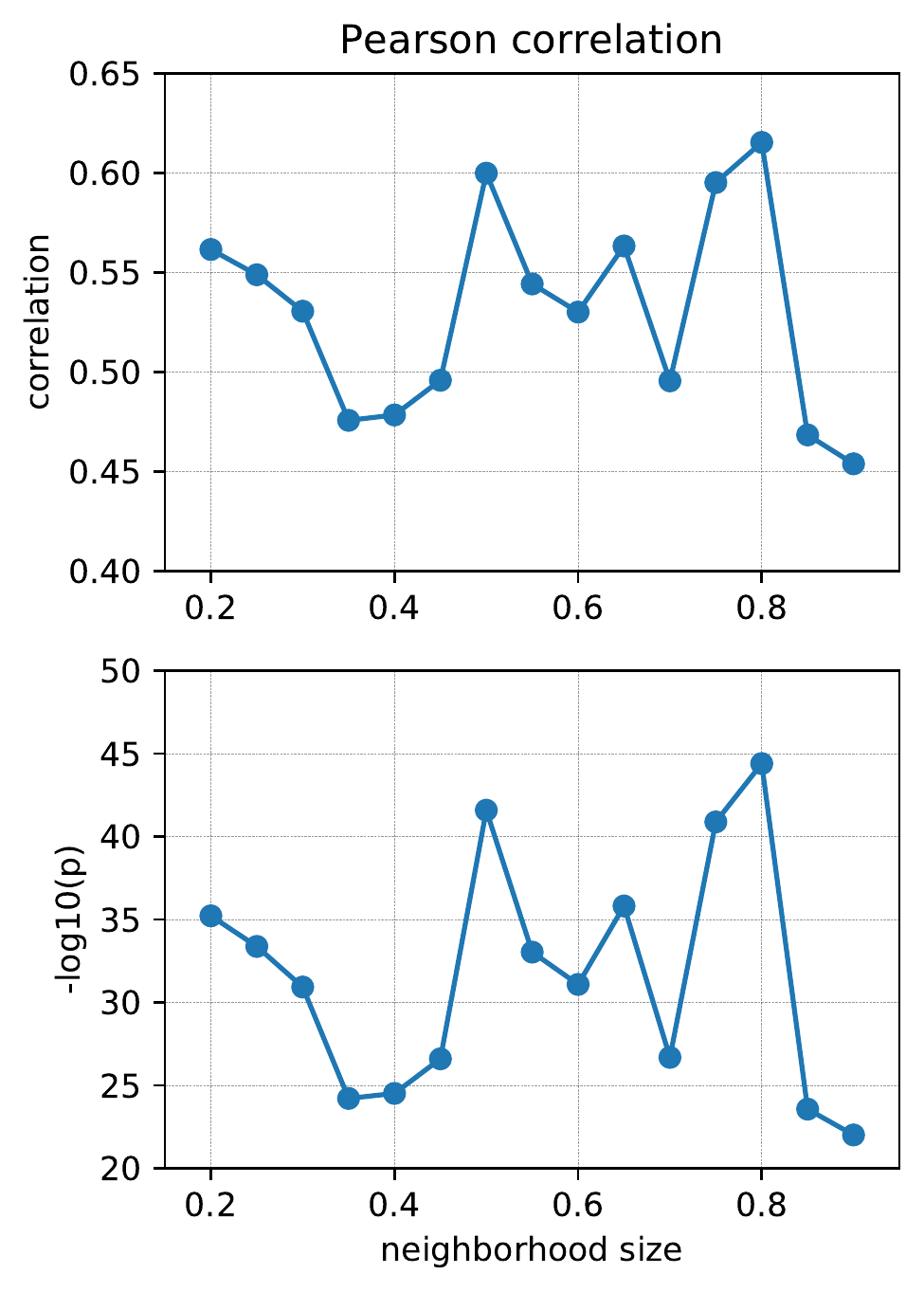}
		\captionsetup{width=0.90\columnwidth}
		\caption{Pearson correlation and p-values between pattern strength and robustness for different neighborhood sizes $t$.}
		\label{fig:real-zsc-r-pearson}
    \end{minipage}
\end{figure*}

\end{document}